\documentclass[12pt]{article}

\pdfoutput=1

\makeatletter
\newif\ifkp@upRm
\DeclareSymbolFont{Letters}{OML}{jkp}{m}{n}
\DeclareMathSymbol{\partialup}{\mathord}{Letters}{128}
\DeclareMathSymbol{\DD}{D}{Letters}{128}
\makeatother

\usepackage{amsmath}
\usepackage{amssymb}
\usepackage{graphicx}
\usepackage[dvipsnames]{xcolor}
\usepackage{soul}
\numberwithin{equation}{section}
\usepackage{array}
\usepackage{mathtools}
\usepackage{dsfont}
\usepackage{mathrsfs}

\usepackage{BOONDOX-uprscr}

\usepackage{tikz}\usetikzlibrary{matrix,fit}
\usepackage{tikz-cd} 
\usepackage{varwidth}
\usepackage{enumerate}
\usepackage{appendix}
\usepackage{xfrac}
\usepackage{nicefrac}
\usepackage{mathtools,slashed}

\usepackage[margin=1in]{geometry}
\usepackage{nicematrix}

\usepackage[
    backend=bibtex,
    style=alphabetic,
maxbibnames=99,
giveninits=true,
minalphanames=1,
maxalphanames=3,url=false
  ]{biblatex}

\bibliography{Tdual.bib}

\usepackage{mathabx}
\usepackage{empheq}

\usepackage{setspace}

\usepackage{slashed}
\usepackage{upgreek}
\usepackage{appendix}

\usepackage{tabstackengine}

\usepackage{wrapfig}
\usepackage[abs]{overpic}

\usepackage{float}

\fixTABwidth{T}

\newdimen\mytextwidth
\newcommand\rem[2][cyan!40!green]{\noindent\nobreak\hfil\penalty1000\hfilneg
\mytextwidth=\linewidth\advance\mytextwidth by 2mm
\begin{tikzpicture}[baseline=-\the\dimexpr\fontdimen22\textfont2\relax]\node[outer sep=0pt,draw=black,fill=#1,fill opacity=1,text opacity=1,rectangle,rounded corners]{\begin{varwidth}{\mytextwidth}\textcolor{white}{#2}\end{varwidth}};
\end{tikzpicture}\allowbreak
}

\newcommand\whiterem[2][white!]{\noindent\nobreak\hfil\penalty1000\hfilneg
\mytextwidth=\linewidth\advance\mytextwidth by 2mm
\begin{tikzpicture}[baseline=-\the\dimexpr\fontdimen22\textfont2\relax]\node[outer sep=0pt,draw=black,fill=#1,fill opacity=1,text opacity=1,rectangle,rounded corners,line width=1.5pt]{\begin{varwidth}{\mytextwidth}\textcolor{black}{#2}\end{varwidth}};
\end{tikzpicture}\allowbreak
}

\makeatletter
\newsavebox{\@brx}
\newcommand{\llangle}[1][]{\savebox{\@brx}{\(\m@th{#1\langle}\)}%
  \mathopen{\copy\@brx\kern-0.5\wd\@brx\usebox{\@brx}}}
\newcommand{\rrangle}[1][]{\savebox{\@brx}{\(\m@th{#1\rangle}\)}%
  \mathclose{\copy\@brx\kern-0.5\wd\@brx\usebox{\@brx}}}
\makeatother

\newcommand{\dd}{\partialup}

\newcommand{\CP}{\mathbb{CP}}
\newcommand{\CC}{\mathbb{C}}

\renewcommand{\tilde}{\widetilde}

\newcommand{\bea}{\begin{equation}}
\newcommand{\eea}{\end{equation}}
\newcommand{\bear}{\begin{eqnarray}}
\newcommand{\eear}{\end{eqnarray}}
\newcommand{\bearr}{\begin{eqnarray*}}
\newcommand{\eearr}{\end{eqnarray*}}

\usepackage{ytableau}
\ytableausetup{centertableaux}

\usepackage{tikz}
\usepackage{xparse}
\NewDocumentCommand{\xrightarrows}{ O{}O{} }{%
\mathrel{%
\vcenter{\hbox{%
\begin{tikzpicture}
  \node[minimum width=1cm,minimum height=1ex,anchor=south,align=center] (a){\text{\vphantom{hg}#1}\\[0.5ex] \vphantom{hg}#2};
  \draw[<-] ([yshift=0.35ex]a.west) -- ([yshift=0.35ex]a.east);
  \draw[->] ([yshift=-0.35ex]a.west) -- ([yshift=-0.35ex]a.east);
\end{tikzpicture}
}}%
}%
}

\renewbibmacro{in:}{}

\usepackage{mdframed}

\newbibmacro{string+doi}[1]{
  \iffieldundef{doi}{#1}{\href{http://dx.doi.org/\thefield{doi}}{#1}}}
\DeclareFieldFormat{title}{\usebibmacro{string+doi}{\mkbibemph{#1}}}
\DeclareFieldFormat[article]{title}{\usebibmacro{string+doi}{\mkbibquote{#1}}}

\newmdenv[
  topline=false,
  bottomline=false,
  rightline=false,
  linewidth=2pt,
  skipabove=\topsep,
  skipbelow=\topsep
]{siderules}

\newmdenv[
  topline=false,
  bottomline=false,
  linewidth=2pt,
  skipabove=\topsep,
  skipbelow=\topsep
]{siderulesright}

\makeatletter
\renewcommand{\@seccntformat}[1]{\csname the#1\endcsname.\quad}
\makeatother

\usepackage{setspace}
\usepackage{xpatch}

\makeatletter
\renewcommand{\@chap@pppage}{
  \clear@ppage
  \thispagestyle{plain}
  \if@twocolumn\onecolumn\@tempswatrue\else\@tempswafalse\fi
  \null\vfil
  \markboth{}{}
  {\centering
   \interlinepenalty \@M
   \normalfont
   \MakeUppercase \appendixpagename\par}
  \if@dotoc@pp
    \addappheadtotoc
  \fi
  \vfil\newpage
  \if@twoside
    \if@openright
      \null
      \thispagestyle{empty}
      \newpage
    \fi
  \fi
  \if@tempswa
    \twocolumn
  \fi
}
\makeatother

\definecolor{navycol}{RGB}{100,150,160}
   \definecolor{pinkcol}{RGB}{242,55,55}
   \definecolor{greencol}{RGB}{50,205,50}

   \definecolor{bluecol}{RGB}{30,144,255}

 \allowdisplaybreaks

\usepackage{titlesec}

\titleformat*{\section}{\large\bfseries}
\titleformat*{\subsection}{\normalsize\bfseries}
\titleformat*{\subsubsection}{\normalsize\bfseries}
\titleformat*{\paragraph}{\large\bfseries}
\titleformat*{\subparagraph}{\large\bfseries}
\titlespacing{\author}{-5pt}{-5pt}{-5pt}[-5pt]

\makeatletter
\renewcommand\subsubsection{\@startsection{subsubsection}{3}{\z@}
                                     {-3.25ex\@plus -1ex \@minus -.2ex}
                                     {-1.5ex \@plus -.2ex}
                                     {\normalfont\normalsize\bfseries}}
\renewcommand\subsection{\@startsection{subsection}{3}{\z@}
                                     {-3.25ex\@plus -1ex \@minus -.2ex}
                                     {-1.5ex \@plus -.2ex}
                                     {\normalfont\normalsize\bfseries}}                                     
\makeatother

\usepackage{pgfplots}
\pgfplotsset{compat=1.15}
\usepackage{mathrsfs}
\usetikzlibrary{arrows}
\usepackage{afterpage}
\usepackage{emptypage}

\DeclareFontFamily{U}{solomos}{}
\DeclareErrorFont{U}{solomos}{m}{n}{10}
\DeclareFontShape{U}{solomos}{m}{n}{
  <-> s*[1.1]  gsolomos8r
}{}

   \usepackage{stmaryrd}

\usepackage{tikz}
\usetikzlibrary{arrows.meta}

\usepackage{indentfirst}

\usepackage{tocloft}
\let \savenumberline \numberline
\def \numberline#1{\savenumberline{#1.}}

\usepackage{etoolbox}
\patchcmd{\tableofcontents}{\@starttoc}{\vspace{-0.3cm}\@starttoc}{}{}

\usepackage{accents}
\newcommand\thickbar[1]{\accentset{\rule{.7em}{.7pt}}{#1}}

\usepackage{hyperref}
\hypersetup{
colorlinks=true,
linkcolor=MidnightBlue,
citecolor=violet,
filecolor=purple,
urlcolor=cyan,
breaklinks=true
}

\renewcommand{\bar}{\thickbar}

\usepackage{upgreek}

\usepackage{amsthm}

\newcounter{Chapcounter}

\newcommand{\chapter}[1] 
{ {\centering          
  \addtocounter{Chapcounter}{1} \Large \underline{\sffamily \texorpdfstring{\textbf{  Chapter \theChapcounter: ~#1}}{Lg}} }   
  \addcontentsline{toc}{section}{ \color{MidnightBlue} \texorpdfstring{Chapter ~}{Lg}\theChapcounter.\texorpdfstring{~~}{Lg} #1 }    
}

\usepackage{scalerel}

\newcommand\reallywidesim[1]{\mathrel{\ThisStyle{%
  \setbox0=\hbox{$\SavedStyle\mkern3mu_{\text{#1}}\mkern3mu$}%
  \mkern1mu\stackengine{1\LMpt}{%
    \stretchto{\scaleto{\SavedStyle\mkern-1mu\sim\mkern-1mu}{.54\wd0}}{1\ht0}%
  }{$\SavedStyle_{\text{#1}}$}{O}{c}{F}{T}{S}\mkern1mu%
}}}

\title{\vspace{-1cm}\textbf{T-duality for toric manifolds \\in $\mathcal{N}=(2, 2)$ superspace} \vspace{1cm}}

\author{ Dmitri Bykov$^{\,a,\,b,\,c,\,d,\,e}$\footnote{Emails:
 bykov@mi-ras.ru, dmitri.v.bykov@gmail.com} \qquad Savva Kutsubin$^{\,f}$\footnote{Email: kucubinsavva@gmail.com} \qquad Andrew Kuzovchikov$^{\,a,\,b,\,c,\,d}$\footnote{Email:
 andrkuzovchikov@mail.ru}
\\  \vspace{0cm}  \\
{\small $a)$ 
\emph{Steklov
Mathematical Institute of Russian Academy of Sciences,}} \\{\small \emph{Gubkina str. 8, 119991 Moscow, Russia} }\\
{\small $b)$ 
\emph{Institute for Theoretical and Mathematical Physics,}} \\{\small \emph{Lomonosov Moscow State University, 119991 Moscow, Russia}}\\
{\small $c)$ \emph{HSE University, 6 Usacheva str., Moscow 119048, Russia}}\\
{\small $d)$ \emph{Moscow Center of Fundamental and Applied Mathematics,}} \\{\small \emph{ Lomonosov Moscow State University, 119991 Moscow, Russia}}\\
{\small $e)$ \emph{Beijing Institute of Mathematical Sciences and Applications (BIMSA),}} \\{\small \emph{Huairou District, Beijing
101408, China}}\\
{\small $f)$ \emph{Moscow Institute of Physics and Technology,}} \\
{\small \emph{Institutskii per. 9, 141702  Dolgoprudny, Russia}}
}

\date{}

\begin{document}

\begin{titlepage}
\centering
{\LARGE\bfseries T-duality for toric manifolds \\ \vspace{0.3cm}\;\,in $\mathcal{N}=(2, 2)$ superspace}

\vspace{1cm}
\renewcommand{\thefootnote}{\fnsymbol{footnote}}
{\large Dmitri Bykov$^{\,a,\,b,\,c,\,d,\,e}$\footnote{Emails:
 bykov@mi-ras.ru, dmitri.v.bykov@gmail.com} \quad Savva Kutsubin$^{\,f}$\footnote{Email: kucubinsavva@gmail.com} \quad Andrew Kuzovchikov$^{\,a,\,b,\,c,\,d}$\footnote{Email:
 andrkuzovchikov@mail.ru}
\vspace{0.5cm}
\newline
{\small $a)$ 
\emph{Steklov
Mathematical Institute of Russian Academy of Sciences,}} \\{\small \emph{Gubkina str. 8, 119991 Moscow, Russia} }\\[4pt]
{\small $b)$ 
\emph{Institute for Theoretical and Mathematical Physics,}} \\{\small \emph{Lomonosov Moscow State University, 119991 Moscow, Russia}}\\[4pt]
{\small $c)$ \emph{HSE University, 6 Usacheva str., Moscow 119048, Russia}}\\[4pt]
{\small $d)$ \emph{Moscow Center of Fundamental and Applied Mathematics,}} \\{\small \emph{ Lomonosov Moscow State University, 119991 Moscow, Russia}}\\[4pt]
{\small $e)$ \emph{Beijing Institute of Mathematical Sciences and Applications (BIMSA),}} \\{\small \emph{Huairou District, Beijing
101408, China}}\\[4pt]
{\small $f)$ \emph{Moscow Institute of Physics and Technology,}} \\[4pt]
{\small \emph{Institutskii per. 9, 141702  Dolgoprudny, Russia}}
} \par
\vfill
\begin{center}
\abstract{ We study the situation when the T-dual of a toric K\"ahler geometry is a generalized K\"ahler geometry involving semi-chiral fields. We explain that this situation is generic for polycylinders, tori and related geometries. Gauging multiple isometries in this case requires the introduction of semi-chiral gauge fields on top of the standard ones. We then apply this technology to the generalized K\"ahler geometry of the $\eta$-deformed $\CP^{n-1}$ model, relating it to the K\"ahler geometry of its T-dual.}
\end{center}
\end{titlepage}

\tableofcontents

\section{Introduction}

The study of supersymmetric (SUSY) sigma models dates back to the work~\cite{WittenCP1, Ferrara} on the $\CP^1$ model in two spacetime dimensions as well as~\cite{Cremmer} (in 4D), \cite{DAdda} (again in 2D)  on the $\CP^{n-1}$ generalization. These results were nicely summarized and generalized in~\cite{Zumino}, where it was pointed out that, in the simplest setting, $\mathcal{N}=(2,2)$ SUSY naturally leads to K\"ahler geometry of the target space. In fact, such models are  reductions of $\mathcal{N}=1$ sigma models in 4D, and moreover, their superspace Lagrangian is exactly the K\"ahler potential of the target space.  

However, it was eventually realized that 2D supersymmetry allows for a broader spectrum of models than those coming from 4D reductions. In particular, there are geometries described by the so-called twisted chiral~\cite{Gates} and semi-chiral~\cite{Buscher_1988} multiplets on top of the standard chiral ones. Concrete examples of sigma models involving such fields were elaborated in~\cite{Schoutens, IvanovKimRocek, SevrinTroost, SevrinWZW}; it was conjectured in~\cite{SevrinTroost} that most general sigma models with $\mathcal{N}=(2,2)$ SUSY could be described in terms of these multiplets. Renormalization properties of such models have been studied in~\cite{GrisaruQuantum, TomasielloGeneralized, HullRocekMonge}.

The full mathematical description of these models was elucidated with the advent of generalized K\"ahler geometry years later~\cite{HitchinGeneralized, Gualtieri, GualtieriKahler} (see~\cite{KoerberReview} for a review). Since then, a lot of work has been done on the matching between the mathematical formulation and superspace techniques, culminating in the work~\cite{LindstromOffShellComplete} where  complete equivalence was established. 

The point most relevant to the present paper has to do with the role of semi-chiral superfields. To understand these, recall an important feature of generalized K\"ahler geometry\footnote{As opposed to usual K\"ahler geometry, generalized K\"ahler geometry features a non-zero torsion. For completeness let us mention that models with complex target space and non-vanishing torsion have also been widely studied in the context of SUSY quantum mechanics, cf.~\cite{IvanovSmilga, Delduc, Fedoruk} or the book~\cite{SmilgaDiffGeom}.}, as well as of general $\mathcal{N}=(2,2)$ sigma models, namely the existence of \emph{two} complex structures. Indeed, one may think of each complex structure as defining the transformation laws of the fields w.r.t. one of the extended left/right supersymmetries. Requiring that these complex structures be compatible with the metric, one arrives at the notion of bi-Hermitean geometry. It was shown in~\cite{Gates} that, in the case that these complex structures commute, the geometry is described by chiral and twisted chiral superfields alone. In turn, semi-chiral multiplets are needed to describe the situation when the two complex structures do not commute. Models involving semi-chiral and anti-semi-chiral superfields have been discussed in~\cite{LindstromSemiChiral, MerrellSemiChiral, Crichigno1, Crichigno2}. Despite the lucid general theory, there has been a shortage of explicit physically relevant examples of models of this type. We hope to fill this gap with the present work.

The second ingredient important for the present paper comes from the subject of integrable deformations of sigma models, in particular the so-called $\eta$-deformation, which has attracted a lot of attention in the past decades; cf.~\cite{HoareIntegr} for a review. Such deformations date back to the work~\cite{Cherednik, FateevSausage, FateevSigma}, but were studied more systematically in~\cite{KlimcikDSADS, KlimcikIntegr1, KlimcikIntegr2} by the example of the principal chiral model. Subsequently this was extended to general symmetric target spaces in~\cite{DelducMagroVicedo}. The main feature of such deformations is that they are believed to preserve integrability of the models. At the classical level this is manifested by the fact that their e.o.m. admit a zero-curvature representation, just like the undeformed models. Another important property, which is assumed to be an avatar of integrability~\cite{FateevSausage, FateevSigma, LukyanovHarmonic}, is that these geometries are stable under renormalization, at least at one loop (i.e. only a finite number of parameters, often one, are subjected to renormalization), cf.~\cite{FateevLitvinov, HoareTseytlinRG1, HoareTseytlinRG2, DelducRGflow, AlfimovLitvinov1,  Lacroix1loop, AlfimovLitvinov2, AlfimovKurakin}. Although~\cite{DelducMagroVicedo} gave a general recipe for constructing deformed models in the case of symmetric target spaces, working out the geometry explicitly is rather challenging in concrete cases. In~\cite{Bykov_2021} an attempt was made to elaborate the geometry of deformed~$\CP^{n-1}$, which is the higher-$n$ version of the $\eta$-deformed~$\CP^1$, the so-called sausage~\cite{FateevSausage}. However, it was realized immediately that there are qualitative differences between $\CP^1$ and the higher-dimensional cases, the main one being that the latter feature a non-topological $B$-field. In~\cite{DemulderGeneralized} it had been suggested that the deformed geometry is generalized K\"ahler, described in general by chiral and semi-chiral multiplets.

Another ingredient that sews together the above two storylines is T-duality (see~\cite{AlvarezGaumeTduality, GiveonTduality} for a review, and~\cite{Cavalcanti} for the mathematical interpretation in the context of generalized geometry). In~\cite{RocekVerlinde} it was shown that, in the simplest setup, T-duality exchanges chiral and twisted chiral fields. This has been generalized to  geometries involving semi-chiral superfields in~\cite{LindstromRocekTdualityGeneral}.  Invariance of Ricci flow under T-duality was studied in~\cite{Haagensen, Streets}. Building on earlier work~\cite{LitvinovSigma} that studied the deformed~$\CP^2$ model, observing that one could eliminate the $B$-field by performing T-duality on both angles, the same strategy was pursued in~\cite{Bykov_2021}, where it was shown that the dual manifold~$(\CP^{n-1}_\eta)^{\vee}$ has a vanishing $B$-field and, moreover, is K\"ahler. An explicit K\"ahler potential was written out.

It is then natural to assume that, starting from the T-dual potential, one may perform the reverse T-duality to obtain the potential of $\CP^{n-1}_\eta$. In standard cases, as described in~\cite{RocekVerlinde}, this amounts to performing the Legendre transform of the potential and replacing the respective chiral fields by twisted chiral ones. Moreover, in the case of a \emph{toric} manifold, we may perform T-duality w.r.t. all variables, so that we expect that all fields become twisted chiral and the geometry remains K\"ahler after T-duality. In such cases the dual variables are the moment maps for the multiple $U(1)$ actions, whereas the dual potential is known as the symplectic potential and has an important geometric meaning, cf.~\cite{Guillemin, Abreu} or~\cite{BykovDelPezzo} for a brief review. However, as mentioned earlier, in the case of the K\"ahler manifold $(\CP^{n-1}_\eta)^\vee$, its dual  $\CP^{n-1}_\eta$ is only generalized K\"ahler and not just K\"ahler. So how is this possible? The key observation that was already made in~\cite{Bykov_2021} is that, in this case, the transformations corresponding to isometries of the metric are not symmetries of the K\"ahler potential. We will explain below, though, that the potential can be made invariant by adding \emph{twisted chiral} fields via a generalized K\"ahler transformation\footnote{We thank M.~Ro\v{c}ek for teaching us the trick of adding auxiliary twisted chiral fields prior to gauging. In a different context this will be further explored in the joint paper~\cite{JointPaper}.} (so that the metric is unaffected by this change). As a result, the `true' isometries will ultimately involve shifts of both chiral and twisted chiral fields, which places one into the realm of generalized K\"ahler geometry. One can then perform T-duality by gauging these isometries; as we shall explain, to this end a new kind of gauge field is needed, one that involves both the ordinary unconstrained real gauge superfield as well as an additional semi-chiral gauge superfield. This example is particularly instructive, because in the limit of vanishing deformation ($\eta \to 0$) the geometry in both T-duality frames becomes K\"ahler again, so one can trace how chiral and twisted chiral fields arise out of the semi-chiral ones in the limit.

In fact, the above phenomenon -- that the K\"ahler potential cannot be made invariant under a symmetry transformation by using just chiral fields -- applies to a much more general class of models than the $\eta$-deformed geometries. The same happens even for flat space $\CC^{n\geq 2}$  when one considers $\geq 2$ generic linear  holomorphic isometries. Since this holds for $\CC^n$, the same will hold for its quotients, for example the polycylinders $\CC^n/\mathbb{Z}^n\simeq (\CC^\ast)^n$, or the tori $\CC^n/\mathbb{Z}^{2n}\simeq \mathbb{T}^{2n}$. This means that, for all of these cases, the T-dual geometry is generalized K\"ahler. The methods developed below allow finding the dual potentials and describing the dual geometry.

\vspace{0.3cm}
The paper is organized as follows. We start in Section~\ref{flat space isometry sec} by recalling that certain isometries might not be symmetries of the K\"ahler potential. We then continue in Section~\ref{gauging sec} by introducing the relevant gauge fields that allow gauging those isometries in certain cases. When the symmetry in question acts on both chiral and twisted chiral fields we propose a new type of gauge superfield that includes a semi-chiral multiplet. In Section~\ref{T-duality sec} we apply this theory to the superspace description of T-duality. A concrete application to the sigma model with target space $\CC^\ast\times \CC^\ast$ is described in detail in Section~\ref{2-torus model sec}. In Section~\ref{deformed cpN sec}  this theory is extended to the case of the $\eta$-deformed $\CP^{n-1}$ sigma model. The appendices cover various technical matters and are referred to in the text. 

\section{The K\"ahler potential and  flat space isometries
}\label{flat space isometry sec}

We start with the simplest sigma model, with target space the complex plane~$\CC$ parametrized by the complex coordinate\footnote{Later on, in our discussion of T-duality, we will compactify certain directions in target space, so that the complex plane will be replaced by a cylinder, for example.
However, the issue of gauging is local and therefore for the moment we will not worry about global identifications.} $\mathbf{Z}$. The K\"ahler potential has the form
\begin{align}\label{CastKahpot}
    \mathcal{K}_{\mathbb{C}}=a\,\mathbf{Z}\bar{\mathbf{Z}}\,,
\end{align}
where $a>0$ is a constant. This leads to the metric $ds^2=a\, d\mathbf{Z}d\bar{\mathbf{Z}}$, admitting an  isometry
\begin{align}\label{Zshift}
    \mathbf{Z} \mapsto \mathbf{Z}+i\,\epsilon\,.
\end{align}There is also, of course, a second translational isometry that we will discuss shortly.

Notice that, as it stands, the K\"ahler potential is not invariant under this transformation (i.e., it is invariant only up to a K\"ahler transformation). It follows from the classic results of~\cite{HullNonlinear} that, in general, this is an obstacle to gauging. We will elaborate the gauging procedure in the next section, and for the moment we will concentrate on situations when the potential can be made invariant. For example, for the case~(\ref{CastKahpot}) one can write an equivalent potential
\begin{align}
    \mathcal{K}_{\mathbb{C}}\simeq{a\over 2}(\mathbf{Z}+\bar{\mathbf{Z}})^2\,,
\end{align}
which is manifestly invariant under~(\ref{Zshift}) and coincides with~(\ref{CastKahpot}) up to a K\"ahler transformation.

The trick of upgrading to a manifestly invariant potential does not always work, though. For example, the metric coming from~(\ref{CastKahpot}) also has a second isometry in the orthogonal direction, corresponding to the shift $\mathbf{Z}\mapsto \mathbf{Z}+\tilde{\epsilon}$. However, there is no way to write a potential invariant under both shifts. Therefore one cannot gauge both isometries simultaneously while preserving the supersymmetry. As we shall now see, there also exist cases when the isometries of a K\"ahler manifold (even $\CC^n$ for ${n\geq 2}$) cannot be gauged purely in the realm of K\"ahler geometry and require passing to the generalized  K\"ahler setup instead. 

As our first non-trivial example  consider the sigma model with target space  $\CC^2=\CC\times \CC$. 
To this end we introduce two complex coordinates $\mathbf{Z}_1, \mathbf{Z}_2$. Using these, we 
write the most general K\"ahler potential that leads to a flat metric: 
\begin{align}\label{KahPotC2}
    \mathcal{K}_{\mathbb{C}^2}=a_{11}\mathbf{Z}_1 \bar{\mathbf{Z}}_1+a_{22} \mathbf{Z}_2 \bar{\mathbf{Z}}_2+a_{12}\mathbf{Z}_1 \bar{\mathbf{Z}}_2+a_{21}\mathbf{Z}_2 \bar{\mathbf{Z}}_1
\end{align}
In the context of supersymmetric models  $\mathbf{Z}_1$ and $\mathbf{Z}_2$ are interpreted as \emph{chiral} fields\footnote{We remind the notation and main concepts of $\mathcal{N}=(2, 2)$ supersymmetry in Appendix~\ref{SUSYapp}.} and  $\mathcal{K}_{\mathbb{C}^2}$ plays the role of $\mathcal{N}=(2, 2)$ superspace Lagrangian.  Throughout the paper we will always take the liberty of denoting complex coordinates on the manifold as well as the corresponding $\mathcal{N}=(2,2)$ superfields (be it chiral, twisted chiral or semi-chiral) with the same bold letters. On top of that, all other $\mathcal{N}=(2,2)$ superfields, not necessarily involving complex coordinates (such as the gauge superfield $\mathbf{V}$), will be denoted by bold letters as well. 

The  metric arising from the potential~(\ref{KahPotC2}) is simply
\begin{align}\label{C2metric}
    ds^2=\sum\limits_{i, j=1}^2\,a_{ij} \,d\mathbf{Z}_i d\bar{\mathbf{Z}}_j
\end{align}
In what follows it will be convenient to parametrize the matrix $\|a_{ij}\|$ as follows:
\begin{align}
a=    \begin{pmatrix}
a_{1} & \upalpha-i\,\upbeta\\ 
\upalpha+i\,\upbeta & a_{2} 
\end{pmatrix}
\end{align}
as parameters $\upalpha$ and $\upbeta$ will play somewhat different roles in our story\footnote{Note that positivity of the metric requires
\begin{align}\label{determinantMetr}
    \mathrm{Det}(a)=a_1a_2-\upalpha^2-\upbeta^2>0
\end{align}
so that, in particular, $\upalpha$ and $\upbeta$ are bounded from above.}. 

We wish to gauge the two isometries of~(\ref{C2metric}) corresponding to the shifts of $\mathbf{Z}_i$'s:
\begin{align}
    \mathbf{Z}_j \mapsto \mathbf{Z}_j+i\,\epsilon_j\,,\quad\quad j=1, 2
\end{align}
Just as in our previous example, the K\"ahler potential is not invariant under these transformations. However, once again one can write an equivalent potential
\begin{align}\label{KahpotC2beforetwisted}
    &\mathcal{K}_{\mathbb{C}^2}\simeq {a_1\over 2}(\mathbf{Z}_1+ \bar{\mathbf{Z}}_1)^2+{a_2\over 2} (\mathbf{Z}_2+ \bar{\mathbf{Z}}_2)^2+\\ \nonumber &+\upalpha\,(\mathbf{Z}_1+\bar{\mathbf{Z}}_1)(\mathbf{Z}_2+ \bar{\mathbf{Z}}_2)+i \,\upbeta\, (\mathbf{Z}_1+\bar{\mathbf{Z}}_1)(\mathbf{Z}_2 -\bar{\mathbf{Z}}_2)
\end{align}
Now the first three terms are invariant, but for $\upbeta\neq 0$ the fourth term is not. This can be cured by adding an auxiliary \emph{twisted chiral} field $\mathbf{S}_2$ as follows:
\begin{align}\label{kahgen4torus}
    &\mathcal{K}_{\mathbb{C}^2}\simeq {a_1\over 2}(\mathbf{Z}_1+ \bar{\mathbf{Z}}_1)^2+{a_2\over 2} (\mathbf{Z}_2+ \bar{\mathbf{Z}}_2)^2+\upalpha\,(\mathbf{Z}_1+\bar{\mathbf{Z}}_1)(\mathbf{Z}_2+ \bar{\mathbf{Z}}_2)+\\ \nonumber &+i \,\upbeta (\mathbf{Z}_1+\bar{\mathbf{Z}}_1)(\mathbf{Z}_2 -\bar{\mathbf{Z}}_2-\mathbf{S}_2+\bar{\mathbf{S}}_2)
\end{align}
Note that the added term $(\mathbf{Z}_1+\bar{\mathbf{Z}}_1)(\mathbf{S}_2-\bar{\mathbf{S}}_2)$ corresponds to a generalized K\"ahler transformation and thus is a total derivative. On the other hand, the new potential is now invariant w.r.t. the combined transformation
\begin{align}\label{ZSshift}
     \mathbf{Z}_j \mapsto \mathbf{Z}_j+i\,\epsilon_j\,,\quad\quad \mathbf{S}_2 \mapsto \mathbf{S}_2+ i\,\epsilon_2
\end{align}
Our goal in the next section will be to learn how to gauge  global symmetries that act simultaneously on chiral and twisted chiral fields. 

\section{Gauging with chiral \& twisted chiral fields}\label{gauging sec}

We start by recalling how symmetries in systems of chiral fields are gauged, subsequently generalizing to the case of symmetries acting simultaneously on chiral and twisted chiral fields.

\subsection{Only chiral fields.} We start with the case of a single chiral field $\mathbf{Z}$ and a K\"ahler potential of the form $\mathcal{F}(\mathbf{Z}+\bar{\mathbf{Z}})$, invariant under $\mathbf{Z} \mapsto \mathbf{Z}+i\,\epsilon$. One can gauge this  isometry\footnote{General aspects of gauging holomorphic isometries on K\"ahler manifolds are discussed in~\cite{HullNonlinear}.} by replacing 
\begin{align}
    \mathbf{Z}+\bar{\mathbf{Z}}\quad \mapsto \quad \mathbf{Z}+\bar{\mathbf{Z}}+\mathbf{V}\,,
\end{align}
where $\mathbf{V}$ is an unconstrained real gauge superfield. The gauge transformations are
\begin{align}
    \mathbf{Z} \mapsto \mathbf{Z}+ \mathbf{\Phi},\quad \bar{\mathbf{Z}}\mapsto \bar{\mathbf{Z}}+\bar{\mathbf{\Phi}}\,,\quad\quad \mathbf{V}\mapsto \mathbf{V}-\mathbf{\Phi}-\bar{\mathbf{\Phi}}\,,
\end{align}
with $\mathbf{\Phi}$ chiral. The obvious gauge invariants w.r.t. such action are
\begin{align}
    \mathbf{\Sigma}:=\bar{\mathcal{D}}_+\mathcal{D}_-\mathbf{V}\,,\quad\quad \bar{\mathbf{\Sigma}}:=\mathcal{D}_+\bar{\mathcal{D}}_-\mathbf{V}
\end{align}
In fact, one can show that $\mathbf{\Sigma}=\bar{\mathbf{\Sigma}}=0$ implies that $\mathbf{V}=\mathbf{\Phi}_0+\bar{\mathbf{\Phi}}_0$, where $\mathbf{\Phi}_0$ is chiral, so that it is pure gauge\footnote{In deriving this, one needs to use the Poincar\'e lemma, i.e. $da=0$ implies $a=db$, so that this is not a purely algebraic fact.}.

\subsection{General case.}\label{general multiplet sec}

In order to introduce gauge fields in a SUSY sytem of chiral and twisted chiral fields\footnote{A proposal in this direction was made in~\cite{LindstromVectorMultiplets, LindstromGeneralizedNonabelian}, and T-duality in this context was discussed in~\cite{LindstromRocekTdualityGeneral}.  Gauging of models with semi-chiral matter fields was also discussed in~\cite{MerrellTduality}. It would be interesting to clarify the relation of these proposals to our method. 
} one may use the formalism of superconnections, reviewed in Appendix~\ref{superconnections app}. Here we will provide the final recipe instead.

Suppose we have a $U(1)$ isometry that acts simultaneously on chiral and twisted chiral fields, such as~(\ref{ZSshift}). To gauge it, on top of the real gauge field $\mathbf{V}$ we introduce an additional semi-chiral gauge field $\mathbf{X}:$ $\bar{\mathcal{D}}_+ \mathbf{X}=0$. The full gauge multiplet will comprise the pair $(\mathbf{V}, \mathbf{X})$. We will postulate the following transformation laws:
\begin{gather}
    \mathbf{Z} \mapsto \mathbf{Z}+\mathbf{\Phi},\quad \mathbf{S} \mapsto \mathbf{S}+\mathbf{\Lambda}\,,\\  \label{VXgaugetransf}\mathbf{V}\mapsto \mathbf{V}-\mathbf{\Phi}-\bar{\mathbf{\Phi}}\,,\quad \mathbf{X}\mapsto \mathbf{X}+\mathbf{\Phi}-\mathbf{\Lambda}\,,
\end{gather}
where $\mathbf{\Phi}$ is chiral and $\mathbf{\Lambda}$ twisted chiral. The combination $\mathbf{Z}-\mathbf{S}-\mathbf{X}$ is therefore gauge-invariant. The Wess-Zumino gauge for the $(\mathbf{V}, \mathbf{X})$ gauge superfield is elaborated in Appendix~\ref{WZ app}.

Out of $\mathbf{V}$ and $\mathbf{X}$ one can construct the second real gauge superfield
\begin{align}\label{Vtcdef}
\tilde{\mathbf{V}}:=\mathbf{V}+\mathbf{X}+\bar{\mathbf{X}}
\end{align}
that shifts as
\begin{align}
\tilde{\mathbf{V}}\mapsto \tilde{\mathbf{V}}-\mathbf{\Lambda}-\bar{\mathbf{\Lambda}}\,,
\end{align}
so that it serves as a gauge field for twisted chiral fields. 

The gauge invariants of the gauge transformations~(\ref{VXgaugetransf}) are
\begin{align}\label{Fgaugeinv}
    \mathbf{F}=\mathcal{D}_-(\mathbf{V}+\mathbf{X})\quad\quad \textrm{and}\quad\quad \bar{\mathbf{F}}=\bar{\mathcal{D}}_-(\mathbf{V}+\bar{\mathbf{X}})
\end{align}
In Appendix~\ref{pure gauge app} we prove that $\mathbf{F}=\bar{\mathbf{F}}=0$ implies that the gauge fields $\mathbf{V}, \mathbf{X}, \bar{\mathbf{X}}$ are pure~gauge.

A comment is in order regarding the FI term of the new gauge field. Clearly, one can add an FI term for the gauge field $\mathbf{V}$ in the usual way. Notice that there is no need to add an additional FI term for $\tilde{\mathbf{V}}$, since $\int\,d^4\theta\, \tilde{\mathbf{V}}\simeq \int\,d^4\theta\, \mathbf{V}$. Besides, an FI term for the $\mathbf{X}$-field, i.e. $ \int\,d^3\theta\,\mathbf{X}$, is also prohibited, as it is Grassmann-odd. Thus, quite generally, the number of FI terms remains the same as for the ordinary real gauge multiplet.

\subsection{Examples.}\label{examples section} As an illustration of our formalism,  let us consider the $U(1)$ quotients for the cases when the symmetry acts purely on chiral superfields versus when it acts on both chiral and twisted chiral fields. We write out component actions for both models in Wess-Zumino gauge in Appendix~\ref{examples app}.

\vspace{0.3cm}

\emph{1)}
In the first case, we have the $U(1)$ acting on a pair of chiral fields via $(\mathbf{Z}_1, \mathbf{Z}_2)\mapsto e^{i\alpha}(\mathbf{Z}_1, \mathbf{Z}_2)$. The corresponding gauged model has the form
\begin{align}\label{L1lagr}
    \mathcal{K}_1=(|\mathbf{Z}_1|^2+|\mathbf{Z}_2|^2)e^{\mathbf{V}}-\zeta \mathbf{V}\,,
\end{align}
with $\zeta$ a real FI term. Extremizing w.r.t. $\mathbf{V}$, one gets $\mathbf{V}=-\log{\left({1\over \zeta}(|\mathbf{Z}_1|^2+|\mathbf{Z}_2|^2)\right)}$. Substituting back in the potential and dropping constants, one finds
\begin{align}
    \mathcal{K}_1\simeq \zeta \log{(|\mathbf{Z}_1|^2+|\mathbf{Z}_2|^2)}\,,
\end{align}
which is the K\"ahler potential of $\CP^1$ written using homogeneous coordinates.

\vspace{0.3cm}
\emph{2)} 
In the second example we will consider a chiral field $\mathbf{Z}$ and a twisted chiral field~$\mathbf{S}$, together with a similar diagonal $U(1)$ action $(\mathbf{Z}, \mathbf{S}) \mapsto e^{i\alpha} (\mathbf{Z}, \mathbf{S})$. The ungauged model that one starts with has the potential $\mathcal{K}_2=|\mathbf{Z}|^2-|\mathbf{S}|^2$ and gives rise to a (positive-definite) flat metric on $\CC^2$. Gauging leads to
\begin{align}
    &\mathcal{K}_2=e^{\mathbf{V}}|\mathbf{Z}|^2-e^{\tilde{\mathbf{V}}}|\mathbf{S}|^2-\zeta\, \mathbf{V}\,
    ,\\
    &\textrm{where}\quad \tilde{\mathbf{V}}=\mathbf{V}+\mathbf{X}+\bar{\mathbf{X}}
\end{align}
and we have added a real FI term $\zeta$.  One can then use the $\mathbf{\Phi}$ and $\mathbf{\Lambda}$ gauge transformations to set $\mathbf{Z}=\mathbf{S}=1$. It remains to extremize w.r.t. $\mathbf{V}$, which gives $\mathbf{V}=-\log{\left({1\over \zeta}(1-e^{\mathbf{X}+\bar{\mathbf{X}}})\right)}$. Setting $\mathbf{U}:=e^{\mathbf{X}}$, one finds
\begin{align}
    \mathcal{K}_2\simeq\zeta \log{\left(1-\mathbf{U}\bar{\mathbf{U}}\right)}
\end{align}
Let us write out the bosonic part of the Lagrangian. Up to a total derivative, it has the form
\begin{align}\label{pu beta gamma system}
    \int\,d^4\theta\,\mathcal{K}_2 \reallywidesim{bos} i{\dd^2\mathcal{K}_2\over \dd \mathbf{U} \dd \bar{\mathbf{U}}}\left(P \dd_+ \bar{U}-\bar{P}\dd_+ U\right)=i\zeta\,\frac{\bar{P}\dd_+ U-P \dd_+\bar{U}}{(1-|U|^2)^2}
\end{align}
This is a first-order action, an example of $\beta\gamma$-system, which is consistent with the results of~\cite{Buscher_1988}, see also~\cite{LindstromBetaGamma}.

\section{T-duality}\label{T-duality sec}

The discussion of gauging above was purely local and knew nothing about the topology of the target space. For the sake of simplicity we therefore assumed that the target space was~$\CC^n$. However, our main goal in this paper is to study T-duality, and gauging is a technical tool necessary to describe it in superspace language. For T-duality to make sense, we will now partially compactify our target spaces, passing from flat space $\CC$ to the cylinder~$\CC^\ast$ by imposing periodicity $\mathbf{Z}\sim \mathbf{Z}+2\pi i$. In the multidimensional case we will analogously impose periodicity $\mathbf{Z}_j\sim \mathbf{Z}_j+2\pi i$ on all our complex coordinates: as a result, from now on our target space is the polycylinder~$(\CC^\ast)^n$.

\subsection{Chiral fields.}

We start with the case of a single complex field $\mathbf{Z}$ subject to the identification $\mathbf{Z}\sim \mathbf{Z}+2\pi i$, with K\"ahler potential $\mathcal{F}(\mathbf{Z}+\bar{\mathbf{Z}})$. 
The relevant  $U(1)$ isometry is  $\mathbf{Z}\mapsto \mathbf{Z}+i\epsilon$.  In order to perform T-duality, one gauges the isometry and adds Lagrange multipliers saying that $\mathbf{V}$ is pure gauge~\cite{RocekVerlinde}:
\begin{align}\label{chiralpotgauged}
    \mathcal{K}=\mathcal{F}(\mathbf{Z}+\bar{\mathbf{Z}}+\mathbf{V})+\mathbf{G}\, \mathbf{\Sigma}+\bar{\mathbf{G}}\, \bar{\mathbf{\Sigma}}\simeq \mathcal{F}(\mathbf{Z}+\bar{\mathbf{Z}}+\mathbf{V})+(\mathbf{T}+\bar{\mathbf{T}}) \,\mathbf{V}
\end{align}
where $\mathbf{T}=\bar{\mathcal{D}}_+\mathcal{D}_-\mathbf{G}$ is twisted chiral. One can now get rid of the chiral fields via the gauge transformation $\mathbf{V}\mapsto \mathbf{V}-\mathbf{Z}-\bar{\mathbf{Z}}$. Finally, one extremizes w.r.t. $\mathbf{V}$, which amounts to a Legendre transform. Expressing $\mathbf{V}$ in terms of $\mathbf{T}+\bar{\mathbf{T}}$, one gets the dual (generalized) K\"ahler potential. 

The original chiral fields remain only in the cross term between $\mathbf{Z}$ and $\mathbf{T}$, which is of the form $-(\mathbf{T}+\bar{\mathbf{T}})(\mathbf{Z}+\bar{\mathbf{Z}})$. At first glance it may seem like a total derivative but in general this term represents a topologically non-trivial closed two-form, which allows finding the resulting periodicity of $\mathbf{T}$, i.e. $\mathbf{T} \sim \mathbf{T} + i/4$.

\subsubsection{Periodicity of $\mathbf{T}$.}\label{Tperiod}
In order to clarify the meaning of the cross-term $(\mathbf{T}+~\bar{\mathbf{T}})~(\mathbf{Z}+~\bar{\mathbf{Z}})$ let us write it out in $\mathcal{N}=(1,1)$ superfields (see Appendix \ref{N=2,2-N=1,1 dictionary} for details). Here and below $\ll\bullet \gg$ means that we are writing the $\mathcal{N}=(2,2)$ expression $\bullet$ in terms of $\mathcal{N}=(1,1)$ superfields:
\begin{align}
    &\ll(\mathbf{T}+\bar{\mathbf{T}})(\mathbf{Z}+\bar{\mathbf{Z}})\gg = \frac{1}{4}\Big[{\color{gray}\mathscr{D}_+\mathscr{D}_-\left(\mathsf{T}+\bar{\mathsf{T}}\right)\cdot\left(\mathsf{Z}+\bar{\mathsf{Z}}\right)-\left(\mathsf{T}+\bar{\mathsf{T}}\right)\cdot\mathscr{D}_+\mathscr{D}_-\left(\mathsf{Z}+\bar{\mathsf{Z}}\right)}+\nonumber\\
    &+\mathscr{D}_+\left(\mathsf{Z}-\bar{\mathsf{Z}}\right)\mathscr{D}_-\left(\mathsf{T}-\bar{\mathsf{T}}\right)-\mathscr{D}_+\left(\mathsf{T}-\bar{\mathsf{T}}\right)\mathscr{D}_-\left(\mathsf{Z}-\bar{\mathsf{Z}}\right)\Big]\,,
\end{align}
where $\mathsf{T}$ and $\mathsf{Z}$ are the $\mathcal{N}=(1,1)$ reductions of $\mathbf{T}$ and $\mathbf{Z}$ respectively. Note that the gray terms give topologically trivial contributions because $\mathsf{T}+\bar{\mathsf{T}}$ and $\mathsf{Z}+\bar{\mathsf{Z}}$ are non-periodic. The last two terms give a non-trivial topological contribution to the action of the sigma model. Essentially we are interested only in the part involving the lowest components $T$ and $Z$ of $\mathsf{T}$ and $\mathsf{Z}$:
\begin{align}
    \mathcal{S}_{\mathrm{top}} = -4\int d^2z \left(\dd_+\theta \,\dd_- \phi\, -\, \dd_+\phi \,\dd_- \theta\right) = -4\int d\theta\wedge d\phi\,,\label{topContr}
\end{align}
where $\theta = \mathrm{Im}\left(Z\right)$ and $\phi = \mathrm{Im}\left(T\right)$. For simplicity let us assume that the worldsheet is a 2-torus $\mathbb{T}^2$. Then, using the Riemann bilinear identity, we arrive at
\begin{align}
    &\mathcal{S}_{\mathrm{top}} = -4\left(\int_A d\theta \int_B d\phi\, - \,\int_A d\phi \int_B d\theta \right) = -8\pi\cdot p\cdot(n\cdot m' -n'\cdot m)\,,\nonumber\\
    &\text{where} \quad\quad \int_B d\phi = p\cdot m'\,,\quad \int_A d\phi = p\cdot n'
\end{align}
Here $A$ and $B$ are the standard homology cycles of $\mathbb{T}^2$, 
$n$ and $m$ are the winding numbers of $\frac{\theta}{2\pi}$ around these cycles, and $p$ is a period of $\phi$ (determining the period of~$\mathbf{T}$). In the path integral we shall sum over all possible values of the winding numbers $n$ and $m$. This leads to the sum
\begin{align}
    \sum_{n} e^{-8\pi i p \cdot n\cdot m'}\sim \sum_{n}\delta(n+4\cdot p \cdot m')\,.
\end{align}
Consequently $p=1/4$, $n', m' \in \mathbb{Z}$ and $\mathbf{T}\sim \mathbf{T}+i/4$.

\subsection{Chiral and twisted chiral fields.}

In the presence of both chiral and twisted chiral superfields the general potential that we wish to dualize has the form
\begin{align}\label{C2potChTwCh}
    \mathcal{F}(\mathbf{Z}+\bar{\mathbf{Z}}, \mathbf{S}+\bar{\mathbf{S}},\mathbf{Z}-\mathbf{S},\bar{\mathbf{Z}}-\bar{\mathbf{S}})
\end{align}
We assume the periodic identification
\begin{align}\label{ZSperiodic}
    \mathbf{Z}\sim \mathbf{Z}+2\pi in\,,\quad \mathbf{S}\sim \mathbf{S}+2\pi in\,,\quad\quad n\in \mathbb{Z}
\end{align}
so that the target space is 
\begin{align}
    \CC^2/\mathbb{Z}\simeq \mathbb{R}^3\times S^1
\end{align}
The action of $U(1)$ along the circle has the form
\begin{align}\label{ZSactionU(1)}
    \mathbf{Z} \mapsto \mathbf{Z}+i\epsilon\,,\quad\quad \mathbf{S} \mapsto \mathbf{S}+i\epsilon\,,
\end{align}
thus leaving the potential~(\ref{C2potChTwCh}) invariant. To perform T-duality, we introduce the gauge fields and require that they be pure gauge:
\begin{align}
    &\mathcal{K}= \mathcal{F}(\mathbf{Z}+\bar{\mathbf{Z}}+\mathbf{V}, \mathbf{S}+\bar{\mathbf{S}}+\tilde{\mathbf{V}},\mathbf{Z}-\mathbf{S}-\mathbf{X}, \bar{\mathbf{Z}}-\bar{\mathbf{S}}-\bar{\mathbf{X}})+\mathbf{G} \,\mathbf{F}+\bar{\mathbf{G}} \,\bar{\mathbf{F}}\simeq\label{Kpot}\\ &\nonumber
    \simeq \mathcal{F}(\mathbf{Z}+\bar{\mathbf{Z}}+\mathbf{V}, \mathbf{S}+\bar{\mathbf{S}}+\tilde{\mathbf{V}},\mathbf{Z}-\mathbf{S}-\mathbf{X}, \bar{\mathbf{Z}}-\bar{\mathbf{S}}-\bar{\mathbf{X}})+\mathbf{Y} (\mathbf{V}+\mathbf{X})+\bar{\mathbf{Y}}(\mathbf{V}+\bar{\mathbf{X}}),
\end{align}
where $\mathbf{Y}:=\mathcal{D}_-\mathbf{G}$ is semi-chiral. Notice that in the case when $\mathcal{F}$ has no dependence on $\mathbf{S}$ and $\bar{\mathbf{S}}$, the fields $\mathbf{X}, \bar{\mathbf{X}}$ only enter via the last two terms, and variation w.r.t. these fields gives $\bar{\mathcal{D}}_+\mathbf{Y}=0$, implying  that $\mathbf{Y}$ is twisted chiral. In this case we identify $\mathbf{Y}\equiv \mathbf{T}$ from before (see~(\ref{chiralpotgauged})).

\subsubsection{Periodicity of $\mathbf{Y}$.}
We will mostly follow Section \ref{Tperiod} to find the periodicity of~$\mathbf{Y}$. We shall start with the following observation: the original Lagrangian corresponding to the potential~(\ref{Kpot}) is not invariant under the isometry
\begin{align}
    \mathbf{Y}\mapsto \mathbf{Y}+i\,\epsilon\,,
\end{align}
which is the counterpart of~(\ref{ZSactionU(1)}) in the T-dual frame. Indeed, it is only invariant up to a total derivative -- qualitatively this is the same effect as in~(\ref{CastKahpot}) or~(\ref{KahpotC2beforetwisted}). Let us add the term $\mathbf{X}'\mathbf{X} + \Bar{\mathbf{X}}'\Bar{\mathbf{X}}$ in order to make $\mathcal{K}$ invariant. Here $\mathbf{X}'$ is semi-chiral, i.e. $\Bar{\mathcal{D}}_+ \mathbf{X}' = 0$, so that the extra terms are locally trivial and do not impose any new constraints on $\mathbf{X}$. By analogy to~(\ref{ZSperiodic}), let us also assume the periodic identification
\begin{align}
    \mathbf{Y}\sim \mathbf{Y}+ipn\,,\quad \mathbf{X}'\sim \mathbf{X}'-ipn\,,\quad n\in \mathbb{Z}\,,
\end{align}
so that $\mathcal{K}$ is invariant, and the constant $p$ is so far unknown. After the transformation $\mathbf{V}\mapsto\mathbf{V}-\mathbf{Z}-\Bar{\mathbf{Z}}\,,\,\mathbf{X}\mapsto\mathbf{X}-\mathbf{S}+\mathbf{Z}$, the original fields $\mathbf{Z}$ and $\mathbf{S}$ remain only in the terms
\begin{align}   \mathbf{Y}\left(\mathbf{S}+\bar{\mathbf{Z}}\right)+\Bar{\mathbf{Y}}\left(\Bar{\mathbf{S}}+\mathbf{Z}\right) + \mathbf{X}'\left(\mathbf{S}-\mathbf{Z}\right)+\Bar{\mathbf{X}}'\left(\Bar{\mathbf{S}}-\Bar{\mathbf{Z}}\right)\,.
\end{align}
The non-trivial topological contribution arises from the imaginary parts of the lowest components of the fields, i.e. $\phi=\mathrm{Im}\left(Y\right), \psi=\mathrm{Im}\left(X'\right)$ and $\theta = \mathrm{Im}\left(S-\Bar{Z}\right)/2 = \mathrm{Im}\left(S+Z\right)/2$, which are angular (periodic) variables: $\phi \sim \theta + pn$, $\psi \sim \psi - pn$ and $\theta \sim \theta + 2\pi m$ for $n,m \in \mathbb{Z}$. Very explicitly, the topological contribution has the form
\begin{align}
    \mathcal{S}_{\mathrm{top}} = -4\int d\theta \wedge d\phi  \,{\color{gray}-4\int d^2 \dd_+\left(\left(\phi + \psi\right)\dd_- \theta\right) }\,.
\end{align}
The last term is trivial because $\phi+\psi$ is non-periodic. Then, using the same machinery as at the end of Section \ref{Tperiod}, we conclude that the period $p$ of $\phi$ is $1/4$, i.e. $\mathbf{Y} \sim \mathbf{Y}+ i/4$.

\section{Application to the $\CC^\ast \times \CC^\ast$ model}\label{2-torus model sec}

We may now apply this theory to the gauging of~(\ref{kahgen4torus}). To this end we introduce the gauge fields $\mathbf{V}_1, \mathbf{V}_2, \mathbf{X}$, where $(\mathbf{V}_2, \mathbf{X})$ is the pair used to gauge the $\epsilon_2$-isometry~(\ref{ZSshift}). Setting all the original fields to zero by a gauge transformation, $\mathbf{Z}_1=\mathbf{Z}_2=\mathbf{S}=0$ (modulo topological terms discussed above), we find the system
\begin{align}
    &\mathcal{K}={a_1\over 2}\mathbf{V}_1^2+{a_2\over 2} \mathbf{V}_2^2+\upalpha\,\mathbf{V}_1\mathbf{V}_2+i \,\upbeta \mathbf{V}_1(\bar{\mathbf{X}}-\mathbf{X})+\label{polycilinderMod}\\ \nonumber &+\mathbf{Y} (\mathbf{V}_2+\mathbf{X})+\bar{\mathbf{Y}}(\mathbf{V}_2+\bar{\mathbf{X}})+\mathbf{V}_1(\mathbf{\Sigma}+\bar{\mathbf{\Sigma}})\,.
\end{align}
Here $\mathbf{\Sigma}$ is the twisted chiral field imposing the constraint that~$\mathbf{V}_1$ is pure gauge. Let us consider the following two cases:

\vspace{0.3cm}\noindent
$\bullet$ $\upbeta=0$. In this case one uses the e.o.m. for $\mathbf{X}$ to deduce that $\mathbf{Y}$ is twisted chiral, so that ultimately one is left with the $\mathbf{V}_1, \mathbf{V}_2$ symmetric model
    \begin{align}\label{Ka12null}
        \mathcal{K}\big|_{\upbeta=0}={a_1\over 2}\mathbf{V}_1^2+{a_2\over 2} \mathbf{V}_2^2+\upalpha\,\mathbf{V}_1\mathbf{V}_2+\mathbf{V}_1\underbrace{(\mathbf{\Sigma}+\bar{\mathbf{\Sigma}})}_{:=\mathbf{s}}+\mathbf{V}_2\underbrace{(\mathbf{Y}+\bar{\mathbf{Y}})}_{:=y}
    \end{align}
    Integrating out the gauge fields, we get
    \begin{align}\label{Ka12nullfin}
        \mathcal{K}\big|_{\upbeta=0} =-{1\over 2d}\,\left(a_1\, \mathbf{y}^2-2\,\upalpha\,\, \mathbf{y}\mathbf{s}+a_2\,\mathbf{s}^2 \right)\,,\quad d=a_1a_2-\upalpha\,^2
    \end{align}

    \vspace{0.3cm}\noindent
    $\bullet$ $\upbeta\neq 0$. In this case we may shift $\mathbf{X}\to {i\over \upbeta}(\mathbf{X}- \mathbf{\Sigma})$ to get rid of $\mathbf{\Sigma}, \bar{\mathbf{\Sigma}}$ altogether (again the cross terms are total  derivatives). In doing so, we assume that the periods of $\mathbf{X}$ and $\mathbf{\Sigma}$ are the same. One then gets
    \begin{align}
        \mathcal{K}={a_1\over 2}\mathbf{V}_1^2+{a_2\over 2} \mathbf{V}_2^2+\upalpha\,\mathbf{V}_1\mathbf{V}_2+\mathbf{V}_1\underbrace{(\mathbf{X}+\bar{\mathbf{X}})}_{:=\mathbf{x}}+\mathbf{V}_2\underbrace{(\mathbf{Y}+\bar{\mathbf{Y}})}_{:=\mathbf{y}}+{i\over \upbeta}(\mathbf{X}\mathbf{Y}-\bar{\mathbf{X}}\bar{\mathbf{Y}})
    \end{align}
    Notice that the quadratic form in $\mathbf{V}_1, \mathbf{V}_2$ is exactly the same as in~(\ref{Ka12null}). Its inversion thus leads to
    \begin{align}\label{Kaa12general}
        \mathcal{K}=-{1\over 2d}\,\left(a_1\, \mathbf{y}^2-2\,\upalpha\,\, \mathbf{x}\mathbf{y}+a_2\,\mathbf{x}^2 \right)+{i\over \upbeta}(\mathbf{X}\mathbf{Y}-\bar{\mathbf{X}}\bar{\mathbf{Y}})
    \end{align}

We see that the end result is a function of the complex coordinates $\mathbf{X}, \mathbf{Y}$ and their conjugates. In Appendix \ref{general torus metric app} we discuss the geometry behind (\ref{Ka12nullfin}) and (\ref{Kaa12general}) and explain in detail how~(\ref{Ka12nullfin}) arises in the limit $\upbeta\to 0$. To put it short, one should study the e.o.m. that the second term in~(\ref{Kaa12general}) leads to in the limit~${\upbeta\to 0}$. Clearly, its variation w.r.t. $\mathbf{X}$ says that $\mathbf{Y}$ is twisted chiral, whereas the variation w.r.t. $\mathbf{Y}$ says that $\mathbf{X}$ is twisted chiral. The remaining generalized K\"ahler potential is then expressed in terms of twisted chiral fields and coincides with~(\ref{Ka12nullfin}).

\section{The deformed $\mathbb{CP}^{n-1}_{\eta}$  model}\label{deformed cpN sec}

Having developed the necessary techniques, we are now in a position to tackle the main case of interest -- that of the so-called $\eta$-deformed $\mathbb{CP}^{n-1}$ geometry (cf.~\cite{DelducMagroVicedo, LitvinovSigma, DemulderGeneralized, Bykov_2021}), which we will denote by $\mathbb{CP}^{n-1}_{\eta}$. 
It was argued in~\cite{DemulderGeneralized} that it is generalized K\"ahler, described in superspace by chiral and semi-chiral superfields\footnote{The relation between various types of superfields to the bi-complex geometry of the target space is elaborated in the reviews~\cite{Lindstr_m_2012, SevrinThompson}.}. On the other hand,   in~\cite{Bykov_2021} it was shown that the \emph{T-dual} of $\mathbb{CP}^{n-1}_{\eta}$, which we shall denote by $\left(\mathbb{CP}^{n-1}_{\eta}\right)^{\vee}$, is an ordinary  K\"ahler geometry. As we demonstrate below, the nature of the phenomenon -- why one metric is K\"ahler, and the dual one generalized K\"ahler, despite the fact that one performs T-duality on all angles -- is essentially the same as that in the polycylinder examples of the previous sections.

We will start with the K\"ahler potential of $\left(\mathbb{CP}^{n-1}_{\eta}\right)^{\vee}$ that was written out explicitly\footnote{As compared to the expression in~\cite{Bykov_2021} here we have made a rescaling $\mathbf{Z}_i\mapsto {\mathbf{Z}_i\over 2\eta}$ and dropped an overall factor of $1\over 2\eta$ in front of the potential.} in~\cite{Bykov_2021}:
\begin{align}\label{rescaled_K}
    &\mathcal{K}^\vee={1\over 2}\sum_{j=2}^{n-1}i\left(\mathbf{Z}_{j}\bar{\mathbf{Z}}_{j-1}-\bar{\mathbf{Z}}_{j}\mathbf{Z}_{j-1}\right)+\sum_{j=1}^{n}\mathcal{P}\left(\mathbf{t}_{j}-\mathbf{t}_{j-1}\right)\,,\\
    &\textrm{where}\quad \mathcal{P}\left(t\right) = i\left(\mathrm{Li}_{2}\left(e^{it}\right)+\frac{t\left(2\pi-t\right)}{4}\right)\quad \textrm{and}\quad \mathbf{t}_{j} = \mathbf{Z}_{j}+\bar{\mathbf{Z}}_{j}
\end{align}
Here the variables are subject to the restriction $\pi\geq \mathbf{t}_j\geq \mathbf{t}_{j-1}\geq -\pi$. Note that $\mathbf{t}_0$ and $\mathbf{t}_n$ are fixed constant parameters that encode the deformation: the deformation vanishes in the limit $\mathbf{t}_0\to \mathbf{t}_n$.  One can thus think of the `deformation parameter' $\eta$ as being related to the difference $\mathbf{t}_n-\mathbf{t}_0$; in the foregoing we will only use $\mathbf{t}_0, \mathbf{t}_n$ without reference to $\eta$.

In~(\ref{rescaled_K}) one also makes the periodic identification $\mathbf{Z}_{j}\sim \mathbf{Z}_{j}+i$ and, just like before, the toric isometries have the form $\mathbf{Z}_{j}\mapsto \mathbf{Z}_{j}+i \,\epsilon_j$. It is thus clear that the non-invariant terms in~(\ref{rescaled_K}) are very similar\footnote{In fact, as it was shown in~\cite{Bykov_2021}, if one formally studies Ricci flow along the imaginary time axis, the metric coming from~(\ref{rescaled_K}) will have the flat polycylinder $(\CC^\ast)^{n-1}$ as its UV limit.} to our example~(\ref{KahPotC2}), so the methods developed earlier can be applied here as well.

\subsection{Gauging isometries.}\label{Gauging isometries.} In order to implement the method proposed in the previous sections, we replace the non-invariant terms in~(\ref{rescaled_K}) by an equivalent potential (for concreteness we shall consider the case of odd $n$ first)
\begin{equation}
\begin{aligned}\nonumber
    & {i\over 2}\sum_{j=1}^{\frac{n-3}{2}}\left(\mathbf{Z}_{2j-1}+\bar{\mathbf{Z}}_{2j-1}-\mathbf{Z}_{2j+1}-\bar{\mathbf{Z}}_{2j+1}\right)\left(\mathbf{Z}_{2j}-\bar{\mathbf{Z}}_{2j}\right)+{i\over 2} \left(\mathbf{Z}_{n-2}+\bar{\mathbf{Z}}_{n-2}\right)\left(\mathbf{Z}_{n-1}-\bar{\mathbf{Z}}_{n-1}\right)
\end{aligned}
\end{equation}
Now we proceed to gauge the isometries  $\mathbf{Z}_{j}\mapsto \mathbf{Z}_{j}+i\epsilon_{j}$. For each $j$ we again introduce an unconstrained real gauge superfield $\mathbf{V}_{j}$, an auxiliary twisted chiral field $\mathbf{S}_{j}$, and a semi-chiral gauge field $\mathbf{X}_{j}$ satisfying  $\bar{\mathcal{D}}_+ \mathbf{X}_{j}=0$. We then make the following replacements: 
\begin{equation}
    \mathbf{Z}_{j}+\bar{\mathbf{Z}}_{j} \mapsto \mathbf{Z}_{j}+\bar{\mathbf{Z}}_{j}+\mathbf{V}_{j}
\end{equation}
\begin{equation}
    \mathbf{Z}_{j}-\bar{\mathbf{Z}}_{j} \mapsto \mathbf{Z}_{j}-\mathbf{S}_{j}-\mathbf{X}_{j}-\bar{\mathbf{Z}}_{j}+\bar{\mathbf{S}}_{j}+\bar{\mathbf{X}}_{j}
\end{equation}
Setting $\mathbf{Z}_{j}=\mathbf{S}_{j}=0$ by a gauge transformation, we obtain:
\begin{equation}
\begin{aligned}
&\mathcal{K} = {i\over 2}\sum_{j=1}^{\frac{n-3}{2}}\left(\mathbf{V}_{2j-1}-\mathbf{V}_{2j+1}\right)\left(\bar{\mathbf{X}}_{2j}-\mathbf{X}_{2j}\right)+{i\over 2} \mathbf{V}_{n-2}\left(\bar{\mathbf{X}}_{n-1}-\mathbf{X}_{n-1}\right) \\& + \sum_{j=1}^{n}\mathcal{P}\left(\mathbf{V}_{j}-\mathbf{V}_{j-1}\right)+\sum_{j=1}^{\frac{n-1}{2}}\left(\mathbf{Y}_{2j}\left(\mathbf{V}_{2j}+\mathbf{X}_{2j}\right)+\bar{\mathbf{Y}}_{2j}\left(\mathbf{V}_{2j}+\bar{\mathbf{X}}_{2j}\right)\right) \\& + \sum_{j=1}^{\frac{n-1}{2}}\mathbf{V}_{2j-1}\left(\mathbf{\Sigma}_{2j-1}+\bar{\mathbf{\Sigma}}_{2j-1}\right)\,,\quad\quad\quad \mathbf{V}_0\equiv \mathbf{t}_0\,,\quad \mathbf{V}_n\equiv \mathbf{t}_n
\end{aligned}
\end{equation}
where we have added Lagrange multipliers indicating that $\mathbf{V}_{j}$ is pure gauge. Notice that the indices of $\mathbf{X}$ and $\mathbf{Y}$ take only even values, while the indices of $\mathbf{\Sigma}$ take odd values. 

It turns out that, just like in our example of Section~\ref{2-torus model sec}, one can eliminate all of the $\mathbf{\Sigma}_{j}, \bar{\mathbf{\Sigma}}_{j}$ fields  by the shift 
\begin{align}\nonumber
    \mathbf{X}_{2m}\mapsto 2i\left(\mathbf{X}_{2m}-\sum_{i=1}^{m}\mathbf{\Sigma}_{2i-1}\right)\,.
\end{align}
As shown in Appendix~\ref{even app}, in the case of even $n$ one can eliminate all twisted chiral fields $\mathbf{\Sigma}_{j}$ but one, the remaining twisted chiral field being\footnote{The fact that, in the case of even $n$, there should be one twisted chiral field on top of the semi-chiral ones, was mentioned in~\cite{DemulderGeneralized}.}  $\mathbf{\Sigma} = \sum_{i=1}^{n/2}\mathbf{\Sigma}_{2i-1}$. 

In order to simplify the above expression we introduce the new variables 
\begin{align}
  &\mathbf{Y}_{2j-1}=\mathbf{X}_{2j}-\mathbf{X}_{2j-2}, \ \mathbf{X}_{0}\equiv \mathbf{X}_{n} \equiv 0\\
  &\mathbf{w}_j=\mathbf{Y}_{j}+\bar{\mathbf{Y}}_{j}\quad \quad j=1, \ldots, n-1
\end{align}
The only difference of the even $n$ case is that one should redefine  $\mathbf{w}_{n-1} = \mathbf{Y}_{n-1}+\bar{\mathbf{Y}}_{n-1}+\mathbf{\Sigma}+\bar{\mathbf{\Sigma}}$. Using these definitions, for any $n$ the K\"ahler potential may be recast as
\begin{equation}\label{VwInhomCoordKahPot}
    \mathcal{K} = \sum_{j=1}^{n-1}\mathbf{V}_{j} \,\mathbf{w}_j+ \sum_{j=1}^{n}\mathcal{P}\left(\mathbf{V}_{j}-\mathbf{V}_{j-1}\right) + 2i\sum_{j=1}^{\lfloor\frac{n-1}{2}\rfloor}\left(\mathbf{Y}_{2j}\mathbf{X}_{2j}-\bar{\mathbf{Y}}_{2j}\bar{\mathbf{X}}_{2j}\right)
\end{equation}

\subsection{The Legendre transform.} In what follows it will be convenient to introduce the new variables 
\begin{align}
    \mathbf{M}_{k}=\mathbf{V}_{k}-\mathbf{V}_{k-1}\quad\quad\quad k=1, \ldots, n
\end{align}
satisfying the constraint 
\begin{align}\label{xsumconstr}
    \sum\limits_{k=1}^n \, \mathbf{M}_k=\mathbf{V}_n-\mathbf{V}_0\equiv \uprho\,,
\end{align}
where $\uprho$ plays the role of deformation parameter. We also define the dual variables 
\begin{align}
    \tilde{\mathbf{w}}_k=\sum\limits_{j=k}^{n-1}\,\mathbf{w}_j\equiv -\log{(\mathbf{U}_k \bar{\mathbf{U}}_k)}\,,\quad\quad k=1,\, \ldots, \,n-1 
\end{align}
together with an additional auxiliary variable $\tilde{\mathbf{w}}_n$. The latter is needed in order to pass  to homogeneous coordinates and will make the final formulas more symmetric.  In this case one may rewrite~(\ref{VwInhomCoordKahPot}) as
\begin{align}
    \mathcal{K} = \sum_{j=1}^{n}\,\mathbf{M}_{j} \,\tilde{\mathbf{w}}_j+ \sum_{j=1}^{n}\mathcal{P}\left(\mathbf{M}_j\right)-\mathbf{L} \left(\sum\limits_{k=1}^n \, \mathbf{M}_k-\uprho\right) + 2i\sum_{j=1}^{\lfloor\frac{n-1}{2}\rfloor}\left(\mathbf{Y}_{2j}\mathbf{X}_{2j}-\bar{\mathbf{Y}}_{2j}\bar{\mathbf{X}}_{2j}\right)\,,\label{kahler potential}
\end{align}
with $\mathbf{L}$ a Lagrange multiplier superfield. As a result of introducing the extra variable, one now has the gauge symmetry ${\tilde{\mathbf{w}}_j \mapsto \tilde{\mathbf{w}}_j+\mathrm{Re}(f(\mathbf{U}))}$, $\mathbf{L} \mapsto \mathbf{L}+\mathrm{Re}(f(\mathbf{U}))$. In order to return to the inhomogeneous coordinates, one can use this gauge symmetry to set $\tilde{\mathbf{w}}_n=0$.

Extremizing~(\ref{kahler potential}) w.r.t. $\mathbf{M}_j$ and $\mathbf{L}$, one arrives at the following system of equations:
\begin{align}
    2\sin{{\mathbf{M}_k\over 2}}=e^{\mathbf{L}} |\mathbf{U}_k|^2\,,\quad\quad k=1, \ldots, n\,,\quad\quad
    \sum\limits_{k=1}^n \, \mathbf{M}_k=\uprho
\end{align}
As a cross-check, let us now take the limit of vanishing  deformation, $\uprho\to 0$. In this case all~$\mathbf{M}_k\to 0$, so that $\sin{{\mathbf{M}_k\over 2}}\simeq {\mathbf{M}_k\over 2}$, and the equations are easily solved as follows:
\begin{align}
    \mathbf{L}=-\log{\left(\sum\limits_{k=1}^n |\mathbf{U}_k|^2\right)}+\textrm{const.}\,,\quad\quad \mathbf{M}_k\simeq \uprho\frac{|\mathbf{U}_k|^2}{\sum\limits_{j=1}^n |\mathbf{U}_j|^2}\,.
\end{align}
Using the expansion $\mathcal{P}(x)={i\pi^2\over 6}+x(\log(x)-1)+\ldots $, one  finds
\begin{align}\label{Kdualundeformed}
    \mathcal{K} = -\uprho\log{\left(\sum\limits_{j=1}^n |\mathbf{U}_j|^2\right)}+\ldots\,,
\end{align}
which is minus the K\"ahler potential of $\CP^{n-1}$. The sign has the following origin: in the limit $\uprho\to 0$ both $(\CP^{n-1})^\vee$ as well as $\CP^{n-1}$ are K\"ahler, but since $(\CP^{n-1})^\vee$ was described by chiral multiplets, the dual geometry is now described by twisted chiral multiplets (see the discussion at the end of Section~\ref{2-torus model sec} on how these emerge from semi-chiral multiplets in the limit). In the latter case the metric of the resulting sigma model is \emph{minus} the second derivative of the potential~\cite{Gates}, so that the potential of the form~(\ref{Kdualundeformed}) gives rise to a positive-definite metric.

\subsubsection{$\CP^2$ case.} 
The explicit description of the geometry arising from (\ref{kahler potential}) seems to be a cumbersome and non-trivial task. Nevertheless, it could be done using the symplectic potential, i.e. the Legendre dual of $\mathcal{K}$, and special coordinates adapted to it. Here we shall demonstrate how this works in the case of $\mathbb{CP}^{2}_{\eta}$, where the (generalized) K\"{a}hler potential\footnote{This case was first considered in~\cite{DemulderGeneralized}, using somewhat different variables.}~(\ref{VwInhomCoordKahPot}) is (upon relabeling $\mathbf{w}_1\equiv \mathbf{x}$ and $\mathbf{w}_2\equiv \mathbf{y}$)
\begin{equation}
\begin{aligned}\label{KFrel}
    & \mathcal{K} = \underbracket{\mathbf{V}_1\mathbf{x}+\mathbf{V}_2\mathbf{y} + \mathcal{P}(\mathbf{V}_3 - \mathbf{V}_2)+ \mathcal{P}(\mathbf{V}_2-\mathbf{V}_1)+\mathcal{P}(\mathbf{V}_1 - \mathbf{V}_0)}_{\equiv \mathrm{F}(\mathbf{x},\mathbf{y})}+\\&\quad\quad  + 2 i (\mathbf{X}\mathbf{Y} - \bar{\mathbf{X}}\bar{\mathbf{Y}})\,,\quad \textrm{where}\quad \mathbf{x} = \mathbf{X}+\bar{\mathbf{X}}\,, \quad \mathbf{y} = \mathbf{Y}+\bar{\mathbf{Y}}
\end{aligned}
\end{equation}
Here we have dropped the indices of $\mathbf{X}$ and $\mathbf{Y}$, and $\mathbf{V}_0, \mathbf{V}_3$  are fixed.

It is useful to pass to the $\mathcal{N}=(1,1)$ description: in terms of the function $\mathrm{F}(\mathbf{x},\mathbf{y})$ the $\mathcal{N}=(1,1)$ Lagrangian has the form
\begin{gather}\nonumber
    \mathcal{K}_{(1,1)} = -\frac{1}{4}\frac{\dd^2 \mathrm{F}}{\dd \mathsf{x} \dd \mathsf{y}}\mathscr{D_+}\left(\frac{\dd \mathrm{F}}{\dd \mathsf{x}}\right)\mathscr{D_-}\left(\frac{\dd \mathrm{F}}{\dd \mathsf{y} }\right)-\frac{1}{4}\mathscr{D}_+ \mathsf{y} \mathscr{D}_-\left(\frac{\dd \mathrm{F}}{\dd \mathsf{y}}\right)-\frac{1}{4}\mathscr{D}_+\left(\frac{\dd \mathrm{F}}{\dd \mathsf{x}}\right) \mathscr{D}_- \mathsf{x} +\\
    +\frac{1}{2}\frac{\dd^2 \mathrm{F}}{\dd \mathsf{x}^2}\mathscr{D}_+\mathsf{\Phi}\mathscr{D}_-\left(\frac{\dd \mathrm{F}}{\dd \mathsf{y}}\right)+\frac{1}{2}\frac{\dd^2 \mathrm{F}}{\dd \mathsf{y}^2}\mathscr{D}_+\left(\frac{\dd \mathrm{F}}{\dd \mathsf{x}}\right)\mathscr{D}_-\mathsf{\Omega}+\nonumber\\+\frac{1}{2}\frac{\dd^2 \mathrm{F}}{\dd \mathsf{x}\dd\mathsf{y}}\left(\mathscr{D}_+\mathsf{\Omega}\mathscr{D}_-\left(\frac{\dd \mathrm{F}}{\dd \mathsf{y}}\right)+\mathscr{D}_+\left(\frac{\dd \mathrm{F}}{\dd \mathsf{x}}\right)\mathscr{D}_-\mathsf{\Phi}\right)-\\
    -\frac{\dd^2 \mathrm{F}}{\dd \mathsf{x}^2}\mathscr{D}_+\mathsf{\Phi}\mathscr{D}_-\mathsf{\Phi}-\frac{\dd^2 \mathrm{F}}{\dd \mathsf{y}^2}\mathscr{D}_+\mathsf{\Omega}\mathscr{D}_-\mathsf{\Omega}-\frac{\dd^2 \mathrm{F}}{\dd \mathsf{x}\dd\mathsf{y}}\left(\mathscr{D}_+\mathsf{\Omega}\mathscr{D}_-\mathsf{\Phi}+\mathscr{D}_+\mathsf{\Phi}\mathscr{D}_-\mathsf{\Omega}\right)+\nonumber\\
    +{\color{gray}\frac{i}{2 }\left(\mathscr{D}_+ \mathsf{X} \mathscr{D}_- \mathsf{Y}-\mathscr{D}_+ \mathsf{Y} \mathscr{D}_- \mathsf{X}\right)-\frac{i}{2}\left(\mathscr{D}_+ \Bar{\mathsf{X}} \mathscr{D}_- \Bar{\mathsf{Y}}-\mathscr{D}_+ \Bar{\mathsf{Y}} \mathscr{D}_- \Bar{\mathsf{X}}\right)}\,,\nonumber
\end{gather}
where $\mathsf{\Phi}:=(\mathsf{X}-\Bar{\mathsf{X}})/(2i)$ and $\mathsf{\Omega}:=(\mathsf{Y}-\Bar{\mathsf{Y}})/(2i)$, $\mathsf{x}=\mathbf{x}\big|$ and $\mathsf{y}=\mathbf{y}\big|$. As we will see shortly, $\mathsf{\Phi}$ and $\mathsf{\Omega}$ are genuine angular variables with periodicity $2\pi$. The gray terms represent an exact two-form, so they may be dropped. 

Just like in the usual K\"ahler setup (cf.~\cite{Guillemin, Abreu}), it is often more illuminating  to express the metric in terms of the dual (symplectic) potential
\begin{align}
    &\mathrm{G}(\mathsf{M},\mathsf{N}):=\mathsf{M}\mathsf{x} + \mathsf{N}\mathsf{y} - \mathrm{F}(\mathsf{x},\mathsf{y})\,,\\
    &\mathrm{G}(\mathsf{M},\mathsf{N}) = -\mathcal{P}(\mathsf{V}_3 - \mathsf{N}) - \mathcal{P}(\mathsf{N} - \mathsf{M})~-\mathcal{P}(\mathsf{M} - \mathsf{V}_0)\,,
\end{align}
where the explicit expression in the second line follows from the  definition of $\mathrm{F}$ in~(\ref{KFrel}). One defines the new coordinates $\mathsf{M}:=\frac{\dd \mathrm{F}}{\dd \mathsf{x}},\mathsf{N}:=\frac{\dd \mathrm{F}}{\dd \mathsf{y}}$, which play the role of (generalized) moment maps for the two $U(1)$ actions; accordingly,  $\mathsf{x} = \frac{\dd \mathrm{G}}{\dd \mathsf{M}}$ and $\mathsf{y} = \frac{\dd \mathrm{G}}{\dd \mathsf{N}}$. We are now in a position to rewrite the Lagrangian more explicitly as follows\footnote{We use the well-known property of the Legendre transform, that is $\mathrm{Hess}(\mathrm{F})=\mathrm{Hess}(\mathrm{G})^{-1}$, where $\mathrm{Hess}(\bullet)$ is the matrix of second derivatives.}:
\begin{gather}\label{K_CP2}
    \mathcal{K}_{(1,1)} = \mathrm{A}\left[\frac{1}{8}\mathscr{D}_+\mathsf{M}\mathscr{D}_-\mathsf{M}-\frac{1}{\mathrm{D}}\mathscr{D}_+\mathsf{M}\mathscr{D}_-\mathsf{\Omega}+\frac{2}{\mathrm{D}}\mathscr{D}_+\mathsf{\Omega}\mathscr{D}_-\mathsf{\Omega}\right]+\nonumber\\
    +\mathrm{C}\left[\frac{1}{8}\mathscr{D}_+\mathsf{N}\mathscr{D}_-\mathsf{N}-\frac{1}{\mathrm{D}}\mathscr{D}_+\mathsf{\Phi}\mathscr{D}_-\mathsf{N}+\frac{2}{\mathrm{D}}\mathscr{D}_+\mathsf{\Phi}\mathscr{D}_-\mathsf{\Phi}\right]+\nonumber\\
    +\mathrm{B}\left[\left(\frac{1}{2\mathrm{D}}-\frac{1}{4}\right)\mathscr{D}_+\mathsf{M}\mathscr{D}_-\mathsf{N}-\frac{1}{\mathrm{D}}\left(\mathscr{D}_+\mathsf{\Omega}\mathscr{D}_-\mathsf{N}+\mathscr{D}_+\mathsf{M}\mathscr{D}_-\mathsf{\Phi}\right)\right]+\nonumber\\
    +\frac{2\mathrm{B}}{\mathrm{D}}\left(\mathscr{D}_+\mathsf{\Omega}\mathscr{D}_-\mathsf{\Phi}+\mathscr{D}_+\mathsf{\Phi}\mathscr{D}_-\mathsf{\Omega}\right)\,,\\
    \text{where}\quad \mathrm{A}:=\left[\cot\left(\frac{\mathsf{N}-\mathsf{M}}{2}\right)
    +\cot\left(\frac{\mathsf{M}-\mathsf{V}_0}{2}\right)\right],\quad\mathrm{B}:=\cot\left(\frac{\mathsf{N}-\mathsf{M}}{2}\right)\nonumber\\
    \mathrm{C}:=\left[\cot\left(\frac{\mathsf{N}-\mathsf{M}}{2}\right)
    +\cot\left(\frac{\mathsf{V}_3-\mathsf{N}}{2}\right)\right],\quad \mathrm{D}:=\mathrm{A}\mathrm{C} - \mathrm{B}^2\,.\nonumber
\end{gather}
\begin{center}
\definecolor{qqqqff}{rgb}{0,0,1}
\definecolor{aqaqaq}{rgb}{0.6274509803921569,0.6274509803921569,0.6274509803921569}
\begin{figure}
    \centering 
\begin{tikzpicture}[line cap=round,line join=round,>=triangle 45,x=1cm,y=1cm,scale=0.7]\label{m_plytope} 
\begin{axis}[
    axis lines=middle,
    xmin=-0.3, xmax=3.0,
    ymin=-0.3, ymax=3.0,
    xtick=\empty,
    ytick=\empty,
    xlabel={$\mathsf{M}$}, 
    ylabel={$\mathsf{N}$},
    xlabel style={at={(ticklabel* cs:1.0)}, anchor=west},
    ylabel style={at={(ticklabel* cs:1.0)}, anchor=south},
    width=10cm, height=10cm,
]
\clip(-0.3,-0.3) rectangle (3.0,3.0);
\fill[line width=2pt,color=qqqqff,fill=qqqqff,fill opacity=0.1] (1,1) -- (1,2) -- (2,2) -- cycle;
\draw [line width=1pt,dotted,color=aqaqaq,domain=-0.3:3.0] plot(\x,{(-0--1*\x)/1});
\draw [line width=1pt,dotted,color=aqaqaq] (1,-0.3) -- (1,3.0);
\draw [line width=1pt,dotted,color=aqaqaq,domain=-0.3:3.0] plot(\x,{(--2-0*\x)/1});
\draw [line width=2pt] (1,1)-- (1,2);
\draw [line width=2pt] (1,2)-- (2,2);
\draw [line width=2pt] (1,1)-- (2,2);
\draw [line width=2pt,color=qqqqff] (1,1)-- (1,2);
\draw [line width=2pt,color=qqqqff] (1,2)-- (2,2);
\draw [line width=2pt,color=qqqqff] (2,2)-- (1,1);
\draw [line width=1pt,dash pattern=on 1pt off 1pt] (1,2) circle (0.15cm);
\begin{scriptsize}
\draw[color=aqaqaq] (0.248,0.55) node {$\mathsf{N}=\mathsf{M}$};
\draw[color=aqaqaq] (1.1,0.1) node {$\mathsf{V}_0$};
\draw[color=aqaqaq] (0.1,2.1) node {$\mathsf{V}_3$};
\draw[color=qqqqff] (1.5,2.1) node {$\uprho$};
\draw[color=qqqqff] (0.9,1.5) node {$\uprho$};
\end{scriptsize}
\end{axis}
\end{tikzpicture}
\caption{The moment polytope of $\CP^2$. Regularity of the metric at the encircled corner is used to identify the periodicities of the angles.}
    \label{m_plytope}
\end{figure}
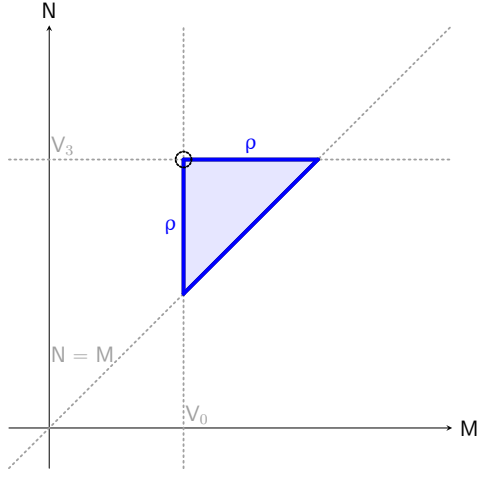
\end{center}

\vspace{-0.5cm}
The moment maps $\mathsf{M}$ and $\mathsf{N}$ take values in the moment polytope. For $\mathbb{CP}^2$, this polytope is a triangle (see Fig.~\ref{m_plytope}). 
To determine the range of the angular variables $\mathsf{\Phi}$ and $\mathsf{\Omega}$ one should consider the limit in which one approaches a vertex of the moment polytope, say $\mathsf{M}\to \mathsf{V}_0$ and $\mathsf{N}\to \mathsf{V}_3$. In this limit, one rescales $\mathsf{M}-\mathsf{V}_0\to \epsilon(\mathsf{M}-\mathsf{V}_0)$ and $\mathsf{N}-\mathsf{V}_3\to \epsilon(\mathsf{N}-\mathsf{V}_3)$ and sends $\epsilon\to 0$, which produces

\begin{gather} \label{cornerlimitpot}
    {1\over \epsilon}\mathcal{K}_{(1,1)} = \frac{1}{4}\frac{1}{\mathsf{M}-\mathsf{V}_{0}}\mathscr{D}_+\mathsf{M}\mathscr{D}_-\mathsf{M}+\frac{1}{4}\frac{1}{\mathsf{V}_{3}-\mathsf{N}}\mathscr{D}_+\mathsf{N}\mathscr{D}_-\mathsf{N}+\\ \nonumber +\left(\mathsf{V}_{3}-\mathsf{N}\right)\mathscr{D}_+\mathsf{\Omega}\mathscr{D}_-\mathsf{\Omega}+\left(\mathsf{M}-\mathsf{V}_{0}\right)\mathscr{D}_+\mathsf{\Phi}\mathscr{D}_-\mathsf{\Phi}+\ldots
\end{gather}
The absence of conical defect at the origin dictates that $\mathsf{\Phi}$ and $\mathsf{\Omega}$ have periodicity~$2\pi$. In this case~(\ref{cornerlimitpot}) describes the flat metric on $\CC^2$, as it should.

Another useful limit is where the deformation vanishes, i.e. $\uprho=\mathsf{V}_3-\mathsf{V}_0 \to 0$. This is the limit where the triangle in Fig.~\ref{m_plytope} formally shrinks to zero. In this case one rescales $(\mathsf{V}_0, \mathsf{M}, \mathsf{N}, \mathsf{V}_3)\to \uprho (\mathsf{V}_0, \mathsf{M}, \mathsf{N}, \mathsf{V}_3)$ and lets $\uprho\to 0$. As a result, one gets:
\begin{gather}
    {1\over \uprho}\mathcal{K}_{(1,1)} = \nonumber\frac{\mathsf{N}-\mathsf{V}_{0}}{4\left(\mathsf{N}-\mathsf{M}\right)\left(\mathsf{M}-\mathsf{V}_{0}\right)}\mathscr{D}_+\mathsf{M}\mathscr{D}_-\mathsf{M}+\frac{\mathsf{V}_{3}-\mathsf{M}}{4\left(\mathsf{N}-\mathsf{M}\right)\left(\mathsf{V}_{3}-\mathsf{N}\right)}\mathscr{D}_+\mathsf{N}\mathscr{D}_-\mathsf{N}-\\-\frac{1}{2\left(\mathsf{N}-\mathsf{M}\right)}\mathscr{D}_+\mathsf{M}\mathscr{D}_-\mathsf{N}+\frac{\left(\mathsf{M}-\mathsf{V}_{0}\right)\left(\mathsf{V}_{3}-\mathsf{N}\right)}{\mathsf{V}_{3}-\mathsf{V}_{0}}\left(\mathscr{D}_+\mathsf{\Omega}\mathscr{D}_-\mathsf{\Phi}+\mathscr{D}_+\mathsf{\Phi}\mathscr{D}_-\mathsf{\Omega}\right)+\\ \nonumber+\frac{\left(\mathsf{V}_{3}-\mathsf{M}\right)\left(\mathsf{M}-\mathsf{V}_{0}\right)}{\mathsf{V}_{3}-\mathsf{V}_{0}}\mathscr{D}_+\mathsf{\Phi}\mathscr{D}_-\mathsf{\Phi}+\frac{\left(\mathsf{V}_{3}-\mathsf{N}\right)\left(\mathsf{N}-\mathsf{V}_{0}\right)}{\mathsf{V}_{3}-\mathsf{V}_{0}}\mathscr{D}_+\mathsf{\Omega}\mathscr{D}_-\mathsf{\Omega}+\ldots
\end{gather}
This can be written in terms of the symplectic potential of $\CP^2$:
\begin{equation}
    {1\over \uprho}\mathcal{K}_{(1,1)} = \frac{1}{4}{\dd^2 \mathrm{G}\over \dd\mu_i \dd\mu_j}\mathscr{D}_+\mu_i\mathscr{D}_-\mu_j+\left({\dd^2 \mathrm{G}\over \dd \mu^2}\right)^{-1}_{ij}\mathscr{D}_+\phi_i\mathscr{D}_-\phi_j+\ldots
\end{equation}
where $\mu_{1}=\mathsf{M}, \ \mu_2 = \mathsf{N}, \ \phi_1=\mathsf{\Phi}, \ \phi_2=\mathsf{\Omega}$ and 
\begin{gather}
\mathrm{G} = \left(\mathsf{M}-\mathsf{V}_{0}\right)\log\left(\mathsf{M}-\mathsf{V}_{0}\right)+\left(\mathsf{V}_{3}-\mathsf{N}\right)\log\left(\mathsf{V}_{3}-\mathsf{N}\right)+\left(\mathsf{N}-\mathsf{M}\right)\log\left(\mathsf{N}-\mathsf{M}\right) \\
    \left({\dd^2 G\over \dd\mu_i \dd\mu_j}\right)_{i,j\in\{1,2\}} = \frac{1}{\mathsf{N}-\mathsf{M}}\begin{pmatrix}
        \frac{\mathsf{N}-\mathsf{V}_{0}}{\mathsf{M}-\mathsf{V}_{0}}& -1 \\
        -1 & \frac{\mathsf{V}_{3}-\mathsf{M}}{\mathsf{V}_{3}-\mathsf{N}}
    \end{pmatrix}
\end{gather}
is the symplectic potential of $\CP^2$~\cite{Abreu}.

\section{Conclusion}

In the present paper we  discussed T-duality in the context of toric K\"ahler manifolds. It was shown that  generically toric isometries cannot be promoted to symmetries of the K\"ahler potential without the introduction of additional `compensator' twisted chiral fields apart from the chiral ones. To gauge such symmetries, we introduced enlarged gauge superfields $(\mathbf{V}, \mathbf{X})$ consisting of a usual gauge superfield $\mathbf{V}$ and a semi-chiral field~$\mathbf{X}$. As a result,  in these cases the T-dual geometry (even upon dualization of all angles) is generalized K\"ahler, described in general by semi-chiral as well as twisted chiral fields. 
As we have explained, the phenomenon is rather general and applies to the cases of polycylinders and higher-dimensional tori. We have considered in detail the case when the target space is $(\CC^\ast)^2$, as well as the  example of the $\eta$-deformed projective space~$\CP^{n-1}_\eta$.

\vspace{1cm}
\textbf{Acknowledgments.} Sections 1-4 were supported by the Russian Science Foundation grant № 25-72-10177 (\href{https://rscf.ru/en/project/25-72-10177/}{\emph{https://rscf.ru/en/project/25-72-10177/}}). Sections 5-7 were supported by the Moscow Center of Fundamental and Applied Mathematics of Lomonosov Moscow State University under agreement № 075-15-2025-345. The work of A.K. was supported by the Foundation for the Advancement of Theoretical Physics and Mathematics ``BASIS". We would like to thank  M.~Alfimov, E.~Ivanov, A.~Litvinov, A.~Smilga for useful discussions, as well as U.~Lindstr\"om and M.~Ro\v{c}ek for many helpful comments on the manuscript. D.B. cordially thanks O.~Hul\'ik, U.~Lindstr\"om, M.~Ro\v{c}ek, R.~von~Unge for the collaboration on a joint  project~\cite{JointPaper} and for the inspiring  discussions on 2D supersymmetry.

\appendix
\renewcommand{\thesection}{\Alph{section}}
\renewcommand{\theequation}{\Alph{section}.\arabic{equation}}
\renewcommand{\theHequation}{\Alph{section}.\arabic{equation}}

\section{Superspace notations}\label{SUSYapp}

Throughout the paper we are working in $\mathcal{N}=(2, 2)$ superspace, following the conventions of~\cite[Chapter 12]{MirrorSymmetryBook}.  The superderivatives have the form
\begin{align}
    &\mathcal{D}_{\pm}={\dd\over \dd \theta^{\pm}}-i\,\bar{\theta}^{\pm}\,{\dd \over \dd x^{\pm}}\\
    &\bar{\mathcal{D}}_{\pm}=-{\dd\over \dd \bar{\theta}^{\pm}}+i\,\theta^{\pm}\,{\dd \over \dd x^{\pm}}
\end{align}
One easily sees that
\begin{align}
    \mathcal{D}_{\pm}^2=\bar{\mathcal{D}}_{\pm}^2=0\,,\quad\quad \{ \mathcal{D}_{\pm}, \bar{\mathcal{D}}_{\mp}\}=0\,, \quad\quad\{\mathcal{D}_{\pm}, \bar{\mathcal{D}}_{\pm}\}=2i {\dd \over \dd x^{\pm}}
\end{align}
We define the chiral and twisted chiral fields:
\begin{align}
    &\textrm{Chiral:}\quad\quad \bar{\mathcal{D}}_+ \mathbf{Z}=\bar{\mathcal{D}}_- \mathbf{Z}=0\\
    &\textrm{Twisted chiral:}\quad\quad \bar{\mathcal{D}}_+ \mathbf{S}=\mathcal{D}_- \mathbf{S}=0
\end{align}
We will also be using semi-chiral fields of two types, that we call $\mathbf{X}$ and $\mathbf{Y}$ in the main text. They are defined by the following constraints:
\begin{align}
    \textrm{Semi-chiral:}\quad\quad\bar{\mathcal{D}}_+ \mathbf{X}=0\,,\quad\quad \mathcal{D}_- \mathbf{Y}=0
\end{align}
All of the above constraints on chiral, twisted chiral and semi-chiral fields may be explicitly resolved. Construct the following combinations:
\begin{align}
    y^{\pm}=x^{\pm}-i\theta^{\pm}\bar{\theta}^{\pm}\,,\quad\quad \bar{y}^\pm=x^{\pm}+i\theta^{\pm}\bar{\theta}^{\pm}
\end{align}
Clearly, they satisfy $\bar{\mathcal{D}}_\pm y^\pm=0$ and $\mathcal{D}_\pm \bar{y}^\pm=0$. Then the fields may be decomposed in components as follows:
\begin{align}
    &\mathbf{Z}=Z(y^+, y^-)+\theta^+ \psi_+(y^+, y^-)+\theta^- \psi_-(y^+, y^-)+\theta^+ \theta^- F(y^+, y^-)\label{chiralcComp}\\
    &\mathbf{S}=S(y^+, \bar{y}^-)+\theta^+ \chi_+(y^+, \bar{y}^-)+\bar{\theta}^- \chi_-(y^+, \bar{y}^-)+\theta^+ \bar{\theta}^- T(y^+, \bar{y}^-)\\
    &\mathbf{X}=X(y^+, x^-)+\theta^+ \xi_+(y^+, x^-)+\theta^- \xi_-(y^+, x^-)+\bar{\theta}^- \zeta_-(y^+, x^-)+\\ \nonumber &+\theta^+ \theta^- G(y^+, x^-)+\theta^+ \bar{\theta}^- H(y^+, x^-)+\theta^-\bar{\theta}^- P(y^+, x^-)+\theta^+\theta^-\bar{\theta}^- \lambda(y^+, x^-)\\
    &\mathbf{Y}=Y(x^+, \bar{y}^-)+\theta^+ \rho_+(x^+, \bar{y}^-)+\bar{\theta}^+ \rho_-(x^+, \bar{y}^-)+\bar{\theta}^- \omega_-(x^+, \bar{y}^-)+\\  &+\theta^+ \bar{\theta}^+ M(x^+, \bar{y}^-)+\theta^+ \bar{\theta}^- N(x^+, \bar{y}^-)+\bar{\theta}^+\bar{\theta}^- Q(x^+, \bar{y}^-)+\theta^+\bar{\theta}^+ \bar{\theta}^- \nu(x^+, \bar{y}^-)\label{semichiralComp}
\end{align}

\section{Superconnections}\label{superconnections app}

Our goal in this appendix is to explain the origin of the generalized gauge multiplet $(\mathbf{V}, \mathbf{X})$ introduced in Section~\ref{general multiplet sec}.

To start with, assume we have a K\"ahler potential of a single chiral superfield $\mathbf{Z}$ of the form $\mathcal{K}=\mathcal{K}(\mathbf{Z}+\bar{\mathbf{Z}})$. Clearly it has a global $U(1)$ symmetry $\mathbf{Z}\mapsto \mathbf{Z}+i \epsilon$, which is the one we wish to gauge. The function $\mathcal{K}$ is formally invariant under local shifts $\mathbf{Z}\mapsto \mathbf{Z}+i\, \epsilon(x^0, x^1)$, but the chirality condition $\bar{\mathcal{D}}_{\pm}\mathbf{Z}=0$ is not. Thus we modify this condition by introducing a superconnection:
\begin{align}
    \label{chiralgauged}
    \bar{\mathcal{D}}_{+}\mathbf{Z}-\bar{\mathbf{\mathbf{A}}}_+=0\,,\quad\quad \bar{\mathcal{D}}_{-}\mathbf{Z}-\bar{\mathbf{\mathbf{A}}}_-=0
\end{align}
These constraints are, of course, invariant under local symmetry transformations, provided one shifts the gauge fields accordingly. 

Acting with $\bar{\mathcal{D}}_{+}$ on the first equation, we find that $\bar{\mathcal{D}}_{+}\bar{\mathbf{\mathbf{A}}}_+=0$, so that we may write $\bar{\mathbf{\mathbf{A}}}_+=\bar{\mathcal{D}}_{+}\mathbf{\mathbf{B}}$ for some field  $\mathbf{B}$. Defining $\mathbf{Z}_1:=\mathbf{Z}-\mathbf{B}$, we rewrite~(\ref{chiralgauged}) as
\begin{align}
    &\bar{\mathcal{D}}_{+}\mathbf{Z}_1=0\,,\quad\quad \bar{\mathcal{D}}_{-}\mathbf{Z}_1-\bar{\mathbf{A}}_-^1=0\\
    &\textrm{where}\quad\quad
    \bar{\mathbf{A}}_-^1=\bar{\mathbf{A}}_--\bar{\mathcal{D}}_- \mathbf{B}
 \end{align}
Again, acting with $\bar{\mathcal{D}}_{-}$ on the second equation we find  $\bar{\mathcal{D}}_{-}\bar{\mathbf{A}}^1_-=0$. Similarly, acting with $\bar{\mathcal{D}}_{+}$ and using the first equation, we find $\bar{\mathcal{D}}_{+}\bar{\mathbf{A}}^1_-=0$. As a result, $\bar{\mathbf{A}}^1_-$ is a chiral field and therefore may be written as $\bar{\mathbf{A}}^1_-=\bar{\mathcal{D}}_{-}\bar{\mathcal{D}}_{+}\mathbf{C}$. Finally, we find that $\mathbf{Z}_2:=\mathbf{Z}_1-\bar{\mathcal{D}}_{+}\mathbf{C}$ is a chiral superfield. Its relation to the original field is
\begin{align}
    \mathbf{Z}=\mathbf{Z}_2+\mathbf{B}+\bar{\mathcal{D}}_{+}\mathbf{C}
\end{align}
In fact, we may simply shift $\mathbf{B}\mapsto \mathbf{B}-\bar{\mathcal{D}}_{+}\mathbf{C}$ to get rid of $\mathbf{C}$ altogether. In this case the superspace Lagrangian takes the form
\begin{align}
    \mathcal{K}=\mathcal{K}(\mathbf{Z}_2+\bar{\mathbf{Z}}_2+\underbrace{\mathbf{B}+\bar{\mathbf{B}}}_{=\mathbf{V}})\,,
\end{align}
where now $\mathbf{Z}_2$ is a standard chiral field, and we have introduced the real gauge superfield $\mathbf{V}$ in order to return to the familiar form of the Lagrangian. The original superconnection is then expressed as
\begin{align}
    \bar{\mathbf{A}}_\pm=\bar{\mathcal{D}}_\pm \mathbf{B}\,.
\end{align}

Let us now discuss what happens in the presence of twisted chiral fields. Clearly, the same construction could be repeated for twisted chiral fields, and the modified twisted chirality condition takes the form
\begin{align}
    \bar{\mathcal{D}}_{+}\mathbf{S}-\bar{\mathbf{A}}'_+=0\,,\quad\quad \mathcal{D}_{-}\mathbf{S}-\mathbf{A}'_-=0
\end{align}
Following the above algorithm results in  
\begin{align}
    \bar{\mathbf{A}}'_+=\bar{\mathcal{D}}_+ \mathbf{B}'\,,\quad\quad \mathbf{A}'_-={\mathcal{D}}_- \mathbf{B}'\,.
\end{align}
If $\mathbf{B}'$ were completely independent of $\mathbf{B}$, this would mean that we are gauging a second $U(1)$ symmetry, independent of the one that acts on the chiral fields. Suppose, instead, that we wish to gauge the same isometry (recall~(\ref{ZSactionU(1)})). Then it is natural to assume that $\bar{\mathbf{A}}'_+=\bar{\mathbf{A}}_+$, so that $\mathbf{B}'=\mathbf{B}+\bar{\mathcal{D}}_+ \mathbf{C}'$. As a result, we have two different \emph{real} gauge superfields $\mathbf{V}$ and~$\tilde{\mathbf{V}}$:
\begin{align}
    &\mathbf{V}=\mathbf{B}+
    \textrm{c.c.}\\
    &\tilde{\mathbf{V}}=\mathbf{B}+\bar{\mathcal{D}}_+\mathbf{C}'+\textrm{c.c.}
\end{align}
In other words,
\begin{align}
    \tilde{\mathbf{V}}-\mathbf{V}=\mathbf{X}+\bar{\mathbf{X}}\,,
\end{align}
where $\mathbf{X}\equiv \bar{\mathcal{D}}_+\mathbf{C}'$ is a semi-chiral superfield satisfying $\bar{\mathcal{D}}_+\mathbf{X}=0$.

\section{${\mathbf{F}=\overline{\mathbf{F}}=0}$ implies the gauge field is pure gauge}\label{pure gauge app}

Here we will show that the vanishing of the gauge field strength $\mathbf{F}=\mathcal{D}_-(\mathbf{V}+\mathbf{X})$ and its complex conjugate $\bar{\mathbf{F}}$ imply that the gauge fields $\mathbf{V}, \mathbf{X}, \bar{\mathbf{X}}$ are pure gauge.

First, it is easy to see that $\mathbf{F}=\bar{\mathbf{F}}=0$ imply $\mathbf{\Sigma}=\bar{\mathbf{\Sigma}}=0$ (since $\mathbf{\Sigma}:=\bar{\mathcal{D}}_+\mathcal{D}_-\mathbf{V}=\bar{\mathcal{D}}_+\mathbf{F}$), so that in this case $\mathbf{V}$ is pure gauge, i.e. 
\begin{align}\label{Vpuregauge}
    \mathbf{V}=\mathbf{\Phi}+\bar{\mathbf{\Phi}}\,
\end{align}
with $\mathbf{\Phi}$ chiral. 
Moreover, $\mathbf{F}=\bar{\mathbf{F}}=0$ also implies $\bar{\mathcal{D}}_+ \bar{\mathcal{D}}_-\tilde{\mathbf{V}}=\mathcal{D}_+ \mathcal{D}_-\tilde{\mathbf{V}}=0$, so that in this case $\tilde{\mathbf{V}}$ is pure gauge as well, i.e. 
\begin{align}\label{Vtpuregauge}
    \tilde{\mathbf{V}}=\mathbf{T}+\bar{\mathbf{T}}\,,
\end{align}
where $\mathbf{T}$ is twisted chiral. Combining $\mathbf{F}=\mathcal{D}_-(\mathbf{V}+\mathbf{X})=0$ with~(\ref{Vpuregauge}), one derives
\begin{align}
    \mathbf{X}=-\mathbf{\Phi}+\mathbf{S}\,,\quad\quad \mathbf{S}=\mathcal{D}_-\mathbf{R}
\end{align}
for some field $\mathbf{R}$. It follows that $\bar{\mathcal{D}}_+\mathbf{S}=\mathcal{D}_-\mathbf{S}=0$, so that $\mathbf{S}$ is twisted chiral. Accordingly, $\bar{\mathbf{X}}=-\bar{\mathbf{\Phi}}+\bar{\mathbf{S}}$. Since $\tilde{\mathbf{V}}=\mathbf{V}+\mathbf{X}+\bar{\mathbf{X}}$, we find
\begin{align}
    \mathbf{T}+\bar{\mathbf{T}}=\mathbf{S}+\bar{\mathbf{S}}\,.
\end{align}
As a result, $i(\mathbf{T}-\mathbf{S})$ is a real twisted chiral superfield. It then easily follows that $i(\mathbf{T}-\mathbf{S})= \alpha$ is a real constant. In other words,
\begin{align}
    \mathbf{X}=-\mathbf{\Phi}+\mathbf{T}+i\,\alpha
\end{align}
The constant may be absorbed by a shift of $\mathbf{\Phi}$. Thus, $\mathbf{X}$ is pure gauge as well.

\section{Wess-Zumino gauge}\label{WZ app}

We will now discuss the suitable version of Wess-Zumino gauge for the new gauge multiplet $(\mathbf{V}, \mathbf{X})$. Since $\mathbf{V}$ is a standard real gauge superfield, one may first bring it to standard Wess-Zumino gauge as follows\footnote{Here $y^{\pm}=x^{\pm}-i\theta^{\pm}\bar{\theta}^{\pm}\,,\; \bar{y}^\pm=x^{\pm}+i\theta^{\pm}\bar{\theta}^{\pm}$, see Appendix~\ref{SUSYapp}.}:
\begin{align}\label{VWZgauge}
    &\mathbf{V}=\theta^+\bar{\theta}^+A_++\theta^-\bar{\theta}^-A_-+\theta^+\bar{\theta}^- \sigma+\theta^-\bar{\theta}^+ \bar{\sigma}+\\ \nonumber &+\theta^+\bar{\theta}^+\theta^-\bar{\lambda}^--\theta^+\bar{\theta}^+\bar{\theta}^-\lambda^-+\theta^-\bar{\theta}^-\theta^+\bar{\lambda}^+-\theta^-\bar{\theta}^-\bar{\theta}^+\lambda^++\theta^+\bar{\theta}^+\theta^-\bar{\theta}^- \mathrm{D} 
\end{align}
The remaining gauge freedom in adding $\mathbf{X}\mapsto \mathbf{X}-\mathbf{\Lambda}$ may be used to bring $\mathbf{X}$ to the following form:
\begin{align}\label{XWZgauge}
    \mathbf{X}=\theta^- \xi_-(y^+, x^-)+\theta^+ \theta^- G(y^+, x^-)+\theta^-\bar{\theta}^- P(y^+, x^-)+\theta^+\theta^-\bar{\theta}^- \lambda(y^+, x^-)
\end{align}

\subsection{Examples.}\phantomsection\label{examples app}

Here we will study the two examples of Section~\ref{examples section} in the Wess-Zumino gauge~(\ref{VWZgauge}).

\vspace{0.3cm}
\emph{1)} 
We start with the more familiar model~(\ref{L1lagr}). In Wess-Zumino gauge the purely bosonic part of the action is (here $\mathbb{D}_\pm$ is the $U(1)$ covariant derivative)
\begin{align}\label{L1lagrWZ}
    \int\,d^4\theta\,\mathcal{K}_1 \reallywidesim{bos} \left(\mathbb{D}_+ Z_1 \,\mathbb{D}_-\bar{Z}_1+
    \mathbb{D}_+ Z_2 \,\mathbb{D}_-\bar{Z}_2+(-\leftrightarrow +)\right)+
    \mathrm{D}\,(|Z_1|^2+|Z_2|^2-\zeta)
\end{align}
The quadratic form of the gauge fields is $ A_+ A_-$, therefore one can integrate out the gauge fields, ultimately arriving at the expression for the metric on $\CP^1$, written in homogeneous coordinates.

\vspace{0.3cm}
\emph{2)} 
To compare with the component action~(\ref{pu beta gamma system}) of the second example we will  study this model in the Wess-Zumino gauge~(\ref{VWZgauge})-(\ref{XWZgauge}) as well. The bosonic part of the Lagrangian reads
\begin{equation}
\begin{aligned}
&\int\,d^4\theta\,\mathcal{K}_2 \reallywidesim{bos} \mathbb{D}_+ Z\, \mathbb{D}_-\bar{Z}+
    \mathbb{D}_- Z \, \mathbb{D}_+\bar{Z}+
    \mathbb{D}_+ S \, \tilde{\mathbb{D}}_-\bar{S}+
    \tilde{\mathbb{D}}_- S \, \mathbb{D}_+\bar{S}+\\& +
    \mathrm{D}\,(|Z|^2-|S|^2+\zeta)-2 S\bar{S}\,\dd_+\mathrm{Im}(P)
    -\sigma \bar{\sigma}(|Z|^2-|S|^2)-S\bar{S}G\bar{G}
\end{aligned}
\end{equation}
Notice that the covariant derivative $\tilde{\mathbb{D}}_-$ acting on the $S$-fields involves the gauge field ${\tilde{A}_-=A_-+\mathrm{Re}(P)}$, which is therefore independent of $A_-$. It is then clear that the quadratic form in $A_+, A_-, \tilde{A}_-$ is degenerate, so that one cannot integrate out the gauge fields. 

Clearly, on the e.o.m. $\sigma=\bar{\sigma}=G=\bar{G}=0$. Let us as well assume $\zeta>0$, so that the constraint $|Z|^2-|S|^2+\zeta=0$ implies $|S|>0$. Using $U(1)$ gauge symmetry, we may require that  $S$ be real positive. In this case one finds
\begin{align}
    \mathbb{D}_+ S \, \tilde{\mathbb{D}}_-\bar{S}+
    \tilde{\mathbb{D}}_- S \, \mathbb{D}_+\bar{S}=2\,\dd_+ S \, \dd_-S+2 S^2\,\tilde{A}_- A_+
\end{align}
The e.o.m. if $\tilde{A}_-$ then implies $A_+=0$, and the Lagrangian takes the form
\begin{equation}
\begin{aligned}
    &\int\,d^4\theta\,\mathcal{K}_2 \reallywidesim{bos} \dd_+ Z \, \dd_-\bar{Z}+
    \dd_- Z \,\dd_+\bar{Z}+
    2\dd_+ S\, \dd_-S+\\&  +iA_-(\bar{Z}\,\dd_+Z-Z \,\dd_+\bar{Z})+
    \mathrm{D}\,(|Z|^2-S^2+\zeta)-2 S^2\,\dd_+\mathrm{Im}(P)
\end{aligned}
\end{equation}
Changing variables $Z=|Z|e^{i\phi}$, using the $\mathrm{D}$-term constraint and integrating by parts, the Lagrangian is recast as
\begin{equation}
\begin{aligned}
    &{1\over 2}\int\,d^4\theta\,\mathcal{K}_2 \reallywidesim{bos} \left(1+{S^2\over |Z|^2}\right)\dd_+ S\,  \dd_-S+
    |Z|^2\dd_- \phi\,  \dd_+\phi+\\& -\,|Z|^2A_- \dd_+\phi+{1\over 2}
    \mathrm{D}\,(|Z|^2-S^2+\zeta)+2S \,\dd_+S\,\mathrm{Im}(P)
\end{aligned}
\end{equation}
By the appropriate shifts of $A_-$ and $\mathrm{Im}(P)$ one can thus remove the first line altogether. In the second line, one can choose $(A_-|Z|^2, \phi)$ and $(S^2, \mathrm{Im}(P))$ as the independent variables. These furnish two pairs of real $\beta\gamma$-systems, thus establishing equivalence with~(\ref{pu beta gamma system}).

\section{$\mathcal{N}=(2,2)\Big{/}\mathcal{N}=(1,1)$ dictionary}\label{N=2,2-N=1,1 dictionary}
The models we are dealing with in the present paper usually involve semi-chiral fields. Their geometric content is not always easy to grasp, as, say, the $\mathcal{N}=(1,1)$ reduction of a \textit{single} semi-chiral field gives rise to \textit{two} $\mathcal{N}=(1,1)$ superfields \cite{Buscher_1988}. However, at the level of $\mathcal{N}=(1,1)$ superfields the geometry of the models is transparent. Thus, in the present appendix we discuss the reduction to $\mathcal{N}=(1,1)$ superspace.

The action of a SUSY sigma model with $\mathcal{N}=(2, 2)$ potential $\mathcal{K}$ is written as follows:
\begin{align}
    &\mathcal{S} = \int d^2 z\; \frac{\dd^2}{\dd \theta^+ \dd \bar\theta^+ } \frac{\dd^2}{\dd \theta^- \dd \bar\theta^- }\mathcal{K}\Big|_{\theta^{\pm}=\bar\theta^{\pm}=0} = \frac{1}{4}\int d^2 z\; \mathscr{D}_+ \mathscr{D}_- Q_+ Q_- \mathcal{K}\Big|_{\theta^{\pm}=\bar\theta^{\pm}=0}\,,\\
    &\text{where} \; \mathscr{D}_{\pm} \equiv \mathcal{D}_{\pm} + \bar{\mathcal{D}}_{\pm}\,,\;Q_{\pm} \equiv i\left(\mathcal{D}_{\pm} - \bar{\mathcal{D}}_{\pm}\right)
\end{align}
are the $\mathcal{N}=(1,1)$ superderivatives and the extra supercharges respectively. The $\mathcal{N}=(1,1)$ Lagrangian is simply $\mathcal{K}_{(1,1)} \equiv1/4 \cdot Q_{+}Q_{-}\mathcal{K}\big|$, where $\big|$ means that we set to zero the `imaginary' parts of the $\theta$'s. 

Let us now summarize the $\mathcal{N}=(1,1)$ superfield content of various  $\mathcal{N}=(2,2)$ superfields:
\begin{align}
    \nonumber
&\mathbf{X}\big|=\mathsf{X}\,,\quad && Q_+ \mathbf{X}\big| = i\mathscr{D}_{+}\mathsf{X}\,, \quad && Q_- \mathbf{X}\big| = \mathsf{\Psi}\,,\quad &&Q_+ Q_- \mathbf{X}\big| = i\mathscr{D}_+ \mathsf{\Psi}\,, \\ 
    &\mathbf{Y}\big|= \mathsf{Y}\,, && Q_+ \mathbf{Y}\big|= \mathsf{\Gamma}\,,\quad &&Q_- \mathbf{Y}\big| = -i\mathscr{D}_{-}\mathsf{Y}\,,\quad &&Q_+ Q_- \mathbf{Y}\big| = i\mathscr{D}_- \mathsf{\Gamma}\,, \\ \nonumber
    &\mathbf{Z} \big| =\mathsf{Z}\,,\quad &&Q_+ \mathbf{Z} \big| = i\mathscr{D}_+ \mathsf{Z}\,,\quad &&Q_- \mathbf{Z} \big| = i\mathscr{D}_- \mathsf{Z}\,,\quad &&Q_+ Q_- \mathbf{Z} \big| = -\mathscr{D}_+\mathscr{D}_- \mathsf{Z}\,,\\ \nonumber
    &\mathbf{S} \big| =\mathsf{S}\,,\quad &&Q_+ \mathbf{S} \big| = i\mathscr{D}_+ \mathsf{S}\,,\quad &&Q_- \mathbf{S} \big| = -i\mathscr{D}_- \mathsf{S}\,,\quad &&Q_+ Q_- \mathbf{S} \big| = \mathscr{D}_+\mathscr{D}_- \mathsf{S}\,,
\end{align}
where we are using the notation of Appendix \ref{SUSYapp}. Note that each semi-chiral field produces two $\mathcal{N}=(1,1)$ superfields: one bosonic and one fermionic. One may worry that we end up with too many fields but, as we shall see, the fermionic fields are typically auxiliary.

\section{The $\mathcal{N}=(1,1)$ description for $\CC^\ast\times \CC^\ast$}
\label{general torus metric app}

In the usual setting the T-dual geometry can be immediately extracted from $\mathcal{K}$ (\ref{Kaa12general}). One simply notes that (twisted) chiral fields may be interpreted as coordinates and $\mathcal{K}$ as a (generalized) K\"ahler potential. However in our situation we are dealing with semi-chiral fields. Thus, let us move to the $\mathcal{N}=(1,1)$ superfield description where the geometry manifests itself clearly.

Using the recipe given in Appendix \ref{N=2,2-N=1,1 dictionary} we write out different parts of the Lagrangian~(\ref{Kaa12general}) in terms of $\mathcal{N}=(1,1)$ superfields:
\begin{align}\nonumber
    &\ll \mathbf{y}^2\gg \;= -i \left(\mathscr{D}_- \Bar{\mathsf{Y}} \;\mathsf{\Gamma} +\Bar{\mathsf{\Gamma}} \;\mathscr{D}_- \mathsf{Y}\right)\,,\\ \nonumber
    &\ll \mathbf{x}^2\gg \;= -i \left(\mathscr{D}_+ \Bar{\mathsf{X}}\; \mathsf{\Psi} +\Bar{\mathsf{\Psi}}\; \mathscr{D}_+ \mathsf{X}\right)\,,\\ 
   & \ll \mathbf{x}\mathbf{y} \gg \;= \frac{1}{4}\Big[i\mathscr{D}_+\left(\mathsf{Y}+\Bar{\mathsf{Y}}\right)\left(\Bar{\mathsf{\Psi}} - \mathsf{\Psi}\right)+i\mathscr{D}_-\left(\mathsf{X}+\Bar{\mathsf{X}}\right)\left(\Bar{\mathsf{\Gamma}} - \mathsf{\Gamma}\right)+\;\\ \nonumber
   &\quad\quad\quad\quad\;\;  +\mathscr{D}_+\left(\mathsf{X}-\Bar{\mathsf{X}}\right)\mathscr{D}_-\left(\mathsf{Y}-\Bar{\mathsf{Y}}\right)- \left(\mathsf{\Psi}+\Bar{\mathsf{\Psi}}\right)\left(\mathsf{\Gamma}+\Bar{\mathsf{\Gamma}}\right)\Big]\,,\\ \nonumber
   &\ll i\left(\mathbf{X}\mathbf{Y} - \Bar{\mathbf{X}}\Bar{\mathbf{Y}}\right)\gg = \frac{1}{4}\Big[-i\left(\mathsf{\Psi}+i \mathscr{D}_- \mathsf{X} \right) \left(\mathsf{\Gamma}-i \mathscr{D}_+ \mathsf{Y}\right)+i\left(\Bar{\mathsf{\Psi}}-i \mathscr{D}_-\Bar{\mathsf{X}}\right)\left(\Bar{\mathsf{\Gamma}}+i \mathscr{D}_+ \Bar{\mathsf{Y}}\right)\Big]+\nonumber\\
   &\quad\quad\quad\quad\quad+\color{gray}\frac{i}{4 }\left(\mathscr{D}_+ \mathsf{X} \mathscr{D}_- \mathsf{Y}-\mathscr{D}_+ \mathsf{Y} \mathscr{D}_- \mathsf{X}\right)-\frac{i}{4}\left(\mathscr{D}_+ \Bar{\mathsf{X}} \mathscr{D}_- \Bar{\mathsf{Y}}-\mathscr{D}_+ \Bar{\mathsf{Y}} \mathscr{D}_- \Bar{\mathsf{X}}\right)\nonumber\,.
\end{align}
The two terms shown in gray  constitute a topologically trivial contribution, and  we will omit them in what follows.

As one can see, the fermionic fields $\mathsf{\Psi}$ and $\mathsf{\Gamma}$ are non-dynamical, and one can eliminate them. Before doing so, note that the last term is multiplied by $1\over \upbeta$, so that in the limit $\upbeta \to 0$ the e.o.m. for the fermions  simply are
\begin{align}
    &\mathsf{\Psi}+i \mathscr{D}_- \mathsf{X}=0\,,\quad\quad \mathsf{\Gamma}-i \mathscr{D}_+ \mathsf{Y}=0\\
    &\Bar{\mathsf{\Psi}}-i \mathscr{D}_-\Bar{\mathsf{X}}=0\,,\quad\quad \Bar{\mathsf{\Gamma}}+i \mathscr{D}_+ \Bar{\mathsf{Y}}=0
\end{align}
which is the condition for the fields $\mathbf{X}$ and $\mathbf{Y}$ to be twisted chiral. Once one substitutes these values, the term ${1\over \upbeta} \ll i\left(\mathbf{X}\mathbf{Y} - \Bar{\mathbf{X}}\Bar{\mathbf{Y}}\right)\gg$ turns out to be of order $\upbeta$ and therefore vanishes. In this way one gets back to the standard theory involving only twisted chiral superfields. 

One could now return to the case of general $\upbeta$ and eliminate the fermionic auxiliary fields (recall that $d=a_1a_2-\upalpha\,^2$):
\begin{gather}
   \mathcal{K}_{(1,1)} = \frac{a_1}{2d}\left(\mathscr{D}_+ \bar{\mathsf{Y}} \;\mathscr{D}_- \mathsf{Y}+\mathscr{D}_+ \mathsf{Y} \;\mathscr{D}_- \bar{\mathsf{Y}}\right) + \frac{a_2}{2d}\left(\mathscr{D}_+ \bar{\mathsf{X}} \;\mathscr{D}_- \mathsf{X}+\mathscr{D}_+ \mathsf{X} \;\mathscr{D}_- \bar{\mathsf{X}}\right)+\nonumber\\
   + \frac{\upalpha}{4d}\left[\mathscr{D}_+ \left(\bar{\mathsf{X}}-\mathsf{X}\right) \mathscr{D}_- \left(\bar{\mathsf{Y}}-\mathsf{Y}\right)-\mathscr{D}_+ \left(\bar{\mathsf{Y}}+\mathsf{Y}\right) \mathscr{D}_- \left(\bar{\mathsf{X}}+\mathsf{X}\right)\right] -\label{K11gen}\\
    - \frac{\upalpha}{2d}\left[\mathscr{D}_+ \mathsf{Y} \mathscr{D}_- \bar{\mathsf{X}} + \mathscr{D}_+ \bar{\mathsf{Y}} \mathscr{D}_- \mathsf{X}\right]- \nonumber\\
    -\frac{a_1a_2\upalpha\,\upbeta^2}{d^3}\mathscr{D}_+\left(\Bar{\mathsf{X}}+\mathsf{X}\right)\mathscr{D}_-\left(\Bar{\mathsf{Y}}+\mathsf{Y}\right)-\nonumber
    -\frac{\upalpha\,^3\upbeta^2}{d^3}\mathscr{D}_+\left(\Bar{\mathsf{
    Y
    }}+\mathsf{Y}\right)\mathscr{D}_-\left(\Bar{\mathsf{X}}+\mathsf{X}\right)+\nonumber\\
    +\frac{\upalpha\,^2\upbeta^2}{d^3}\Big[a_2\mathscr{D}_+\left(\Bar{\mathsf{X}}+\mathsf{X}\right)\mathscr{D}_-\left(\Bar{\mathsf{X}}+\mathsf{X}\right)+a_1\mathscr{D}_+\left(\Bar{\mathsf{Y}}+\mathsf{Y}\right)\mathscr{D}_-\left(\Bar{\mathsf{Y}}+\mathsf{Y}\right)\Big]\nonumber\\
    +\frac{i\; \upbeta}{d^2}\Big[a_1a_2\left(\mathscr{D}_+\Bar{\mathsf{X}}\mathscr{D}_-\Bar{\mathsf{Y}}-\mathscr{D}_+\mathsf{X}\mathscr{D}_-\mathsf{Y}\right)+\upalpha\,^2\left(\mathscr{D}_+\Bar{\mathsf{Y}}\mathscr{D}_-\Bar{\mathsf{X}}-\mathscr{D}_+\mathsf{Y}\mathscr{D}_-\mathsf{X}\right)\Big]+ \nonumber\\ \nonumber
    +\frac{i\; \upalpha\,\upbeta}{d^2}\Big[a_2\left(\mathscr{D}_+\mathsf{X}\mathscr{D}_-\mathsf{X}-\mathscr{D}_+\Bar{\mathsf{X}}\mathscr{D}_-\Bar{\mathsf{X}}\right)+a_1\left(\mathscr{D}_+\mathsf{Y}\mathscr{D}_-\mathsf{Y}-\mathscr{D}_+\Bar{\mathsf{Y}}\mathscr{D}_-\Bar{\mathsf{Y}}\right)\Big]\,.
\end{gather}
For general values of $\upalpha\,$ and $\upbeta$ the expression is rather cumbersome.
The result simplifies tremendously for the following two distinguished cases:

\vspace{0.3cm}\noindent
$\bullet$ $\upbeta = 0$. In this case we are in the situation of~(\ref{Ka12null}), and no semi-chiral fields are needed. In $\mathcal{N}=(1, 1)$ superspace the Lagrangian reads

\begin{gather}
   \mathcal{K}_{(1,1)} =\frac{a_1}{2d}\left(\mathscr{D}_+ \bar{\mathsf{Y}} \;\mathscr{D}_- \mathsf{Y}+\mathscr{D}_+ \mathsf{Y} \;\mathscr{D}_- \bar{\mathsf{Y}}\right) + \frac{a_2}{2d}\left(\mathscr{D}_+ \bar{\mathsf{X}} \;\mathscr{D}_- \mathsf{X}+\mathscr{D}_+ \mathsf{X} \;\mathscr{D}_- \bar{\mathsf{X}}\right)+\nonumber\\
   + \frac{\upalpha}{4d}\left[\mathscr{D}_+ \left(\bar{\mathsf{X}}-\mathsf{X}\right) \mathscr{D}_- \left(\bar{\mathsf{Y}}-\mathsf{Y}\right)-\mathscr{D}_+ \left(\bar{\mathsf{Y}}+\mathsf{Y}\right) \mathscr{D}_- \left(\bar{\mathsf{X}}+\mathsf{X}\right)\right] - \label{Im(a12)=0}\\
    - \frac{\upalpha}{2d}\left[\mathscr{D}_+ \mathsf{Y} \mathscr{D}_- \bar{\mathsf{X}} + \mathscr{D}_+ \bar{\mathsf{Y}} \mathscr{D}_- \mathsf{X}\right]\,.\nonumber
\end{gather}
Let us elaborate the bosonic sigma model part of the Lagrangian (\ref{Im(a12)=0}), which will allow us to read off the geometry, i.e. the metric and the $B$-field. To this end we introduce the lower components $x$ and $y$ of $\mathsf{X}$ and $\mathsf{Y}$ respectively. The bosonic part has the following simple form 
\begin{align}
    &\mathcal{L} = \frac{a_1}{d}\left(\dd_{+}w\dd_{-}w + \dd_{+}\psi\dd_{-}\psi\right)+\frac{a_2}{d}\left(\dd_{+}u\dd_{-}u + \dd_{+}\phi\dd_{-}\phi\right)-\nonumber\\&
    -\frac{\upalpha\,}{d}\left(2\dd_{+}w\dd_{-}u + \dd_{+}\phi\dd_{-}\psi+\dd_{+}\psi\dd_{-}\phi\right)\,,\label{metricIm(a12)=0}
\end{align}
where $x = u + i\phi$ and $y = w+i\psi$. Let us emphasize that $\phi$ and $\psi$ are periodic (angular) coordinates. Note that the $B$-field is proportional to $\dd_{+}w\dd_{-}u - \dd_{+}u\dd_{-}w$ and is topologically trivial, since $u$ and $w$ are non-periodic.

\vspace{0.3cm}\noindent
$\bullet$  $\upalpha\, = 0$ (so that $d=a_1a_2$). In this case
\begin{align}
    &\mathcal{K}_{(1,1)} = \frac{1}{2a_2}\left(\mathscr{D}_+ \bar{\mathsf{Y}} \;\mathscr{D}_- \mathsf{Y}+\mathscr{D}_+ \mathsf{Y} \;\mathscr{D}_- \bar{\mathsf{Y}}\right) + \frac{1}{2a_1}\left(\mathscr{D}_+ \bar{\mathsf{X}} \;\mathscr{D}_- \mathsf{X}+\mathscr{D}_+ \mathsf{X} \;\mathscr{D}_- \bar{\mathsf{X}}\right)+\nonumber\\&
    +\frac{i\,\upbeta}{a_1a_2}\left(\mathscr{D}_+ \bar{\mathsf{X}}\;\mathscr{D}_- \bar{\mathsf{Y}}-\mathscr{D}_+ \mathsf{X}\;\mathscr{D}_- \mathsf{Y}\right)\,.\label{Re(a12)=0}
\end{align}
Using the same notation as above, we arrive at the following expression for the bosonic sigma model Lagrangian corresponding to (\ref{Re(a12)=0}):
\begin{align}
    &\mathcal{L} = \frac{1}{a_2}\left(\dd_{+}w\dd_{-}w + \dd_{+}\psi\dd_{-}\psi\right)+\frac{1}{a_1}\left(\dd_{+}u\dd_{-}u + \dd_{+}\phi\dd_{-}\phi\right)+\nonumber\\&+\frac{2\,\upbeta}{a_1a_2}\left(\dd_{+}\phi \dd_{-}w+\dd_{+}u\dd_{-}\psi\right)\,.\label{BosonicRe(a12)=0}
\end{align}

\subsection{$\upbeta \to \infty$ limit.}\phantomsection
Let us now consider another special limit  $\upbeta \to \infty$. From a geometric point of view, it is ill-defined because the original metric seizes to be positive-definite in this limit (see (\ref{determinantMetr})). Despite this, in the T-dual frame the case $\upbeta \to \infty$ provides an interesting example of $\beta\gamma$-system combined with a sigma model  (see~\cite{LindstromBetaGamma} for a discussion of related systems). 

After eliminating the fermionic fields $\mathsf{\Psi}$ and $\mathsf{\Gamma}$ the $\mathcal{N}=(1,1)$ Lagrangian (\ref{K11gen}) diverges in the considered limit, while the original $\mathcal{N}=(2,2)$ Lagrangian (\ref{Kaa12general}) does not. This indicates that the elimination of the fermionic $\mathcal{N}=(1,1)$ superfields is not allowed when $\upbeta \to \infty$. We will now consider two different cases:

\vspace{0.3cm}\noindent
$\bullet$ $\upalpha = 0$. In this case one  cannot eliminate $\mathsf{\Psi}$ and $\mathsf{\Gamma}$. We end up with 
\begin{align}
    \mathcal{K}_{(1,1)} = \frac{i}{2a_1}\left(\mathscr{D}_+ \Bar{\mathsf{X}} \;\mathsf{\Psi} +\Bar{\mathsf{\Psi}} \;\mathscr{D}_+ \mathsf{X}\right) + \frac{i}{2a_2}\left(\mathscr{D}_- \Bar{\mathsf{Y}} \;\mathsf{\Gamma} +\Bar{\mathsf{\Gamma}} \;\mathscr{D}_- \mathsf{Y}\right)\,.
\end{align} 
Clearly, it is an $\mathcal{N}=(1,1)$ SUSY $\beta\gamma$-system.

\vspace{0.3cm}\noindent
$\bullet$ $\upalpha \neq 0$. Let us decompose the relevant fields into real and imaginary parts: $\mathsf{X} =\mathsf{X}_{R}+i\mathsf{X}_I\,, \,\mathsf{Y} =\mathsf{Y}_{R}+i\mathsf{Y}_I$ and $\mathsf{\Psi} =\mathsf{\Psi}_{R}+i\mathsf{\Psi}_I\,, \,\mathsf{\Gamma} =\mathsf{\Gamma}_{R}+i\mathsf{\Gamma}_I$. Then it turns out that one can integrate out  $\mathsf{\Psi}_{R}$ and $\mathsf{\Gamma}_{R}$:
\begin{align}
    &\mathcal{K}_{(1,1)} = \frac{1}{\upalpha}\mathscr{D}_+ \mathsf{X}_I\mathscr{D}_- \mathsf{Y}_I\,+\frac{a_1-\upalpha}{d}\mathsf{\Gamma}_I\mathscr{D}_-\mathsf{W}_+ + \frac{a_2-\upalpha}{d}\mathsf{\Psi}_I\mathscr{D}_+\mathsf{W}_-\,,\\
    &\text{where}\quad \mathsf{W}_\pm = \mathsf{X}_R \pm \mathsf{Y}_R\,.\nonumber
\end{align}

\section{$\CC^\ast\times \CC^\ast$ in the WZ gauge}
In this section we will discuss the polycylinder model (\ref{polycilinderMod}) in the WZ gauge. The goal is to reproduce the standard Buscher rules~\cite{Buscher1, Buscher2}. For simplicity we assume $\upalpha = 0$, in which case the $\mathcal{N}=(2,2)$ Lagrangian of the model is given by 
\begin{align}
    &\mathcal{K}=\frac{a_1}{2}\left(\mathbf{Z}_1+ \Bar{\mathbf{Z}}_1 + \mathbf{V}_1\right)^2+ \frac{a_2}{2}\left(\mathbf{Z}_2+ \Bar{\mathbf{Z}}_2 + \mathbf{V}_2\right)^2 +\mathbf{Y}\left(\mathbf{V}_2+\mathbf{X}\right)+\Bar{\mathbf{Y}}\left(\mathbf{V}_2+\Bar{\mathbf{X}}\right)+\nonumber\\
    &+i\upbeta \left(\mathbf{Z}_1+ \Bar{\mathbf{Z}}_1 + \mathbf{V}_1\right)\left(\mathbf{Z}_2-\mathbf{S}_2 - \mathbf{X}-\Bar{\mathbf{Z}}_2+\Bar{\mathbf{S}}_2 + \Bar{\mathbf{X}}\right)+\mathbf{V}_1\left(\mathbf{\Sigma}+\Bar{\mathbf{\Sigma}}\right)\,.\label{WZgaugePolyAction}
\end{align}

We will now impose the WZ gauge by requiring that the $\mathcal{N}=(2,2)$ fields admit the component decompositions given by~(\ref{chiralcComp})-(\ref{semichiralComp}) and~(\ref{VWZgauge}),~(\ref{XWZgauge}). We wish to write down the purely bosonic part of (\ref{WZgaugePolyAction}) upon elimination of the auxiliary fields. Let $S_2, Z_j,\Sigma,Y$ be the lowest components of the respective superfields, and $s_2 := \mathrm{Re}\left(S_2\right)$, $Z_{j} = u_j + i\alpha_j$ ($j=1, 2$), $\xi := \mathrm{Im}\left(\Sigma\right)$, $\zeta := \mathrm{Im}\left(Y\right)$. The Lagrangian takes the form
\begin{align}
    &\mathcal{L} = a_1 B_+^1 B_-^1 + a_2 B_+^2 B_-^2 + 4\left(a_1-\frac{\upbeta^2}{a_2}\right)\dd_+ u_1 \dd_- u_1 + 4\left(a_2-\frac{\upbeta^2}{a_1}\right)\dd_+ u_2 \dd_- u_2-\nonumber\\
    &-2\left(\zeta-\upbeta u_1\right)\left(\dd_+ A_-^2-\dd_- A_+^2\right)-2\left(\xi - \upbeta s_2\right)\left(\dd_+ A_-^1-\dd_- A_+^1\right)\,,\label{CCinWZGauge}\\
    &\text{where}\quad B_\pm^1= 2\dd_{\pm}\left(\alpha_1+\frac{\upbeta}{a_1}u_2\right)+A_\pm^1\,,\;B_\pm^2= 2\dd_{\pm}\left(\alpha_2-\frac{\upbeta}{a_2}u_1\right)+A_\pm^2\,.\nonumber
\end{align}
We will shift $\zeta \mapsto \zeta +\upbeta u_1$, $\,\xi \mapsto \xi + \upbeta s_2$ to simplify computations and  remove~$s_2$~completely.

The action (\ref{CCinWZGauge}) can be used to derive the Buscher rules in a standard fashion (see \cite{AlvarezGaumeTduality} for details). One ultimately arrives at 
\begin{align}
    &\mathcal{L} = 4\Big[a_1 \dd_+ u_1 \dd_- u_1 + a_2 \dd_+ u_2 \dd_- u_2 +\frac{1}{a_1} \dd_+ (\xi+\upbeta u_2) \dd_- (\xi - \upbeta u_2) + \nonumber\\
    &+\frac{1}{a_2} \dd_+ (\zeta-\upbeta u_1) \dd_- (\zeta+\upbeta u_1)\Big]\,.\label{BuscherRules}
\end{align}
It is easy to see that the Lagrangians~(\ref{BuscherRules}) and (\ref{BosonicRe(a12)=0}) coincide after the redefinitions
\begin{align}
    &u = 2a_1 u_1\,\,,\quad &&\phi = -2(\xi + \upbeta u_2)\,,\\
    &w= 2a_2 u_2\,,\quad && \psi = -2(\zeta+ \upbeta u_1)\,.
\end{align}

\section{On the $\mathbb{CP}^{n-1}_{\eta}$ model for even $n$}\label{even app}
In Section~\ref{Gauging isometries.} we  eliminated all twisted chiral fields by shifting the semi-chiral gauge fields $\mathbf{X}_{j}$. Here we demonstrate that,  for $n$ even, one twisted chiral field will remain. For even $n$ the equivalent form of the potential \eqref{rescaled_K} is

\begin{equation}
\begin{aligned}
    &\mathcal{K}^\vee \simeq {i\over 2}\sum_{j=1}^{\frac{n-2}{2}}\left(\mathbf{Z}_{2j-1}+\bar{\mathbf{Z}}_{2j-1}-\mathbf{Z}_{2j+1}-\bar{\mathbf{Z}}_{2j+1}\right)\left(\mathbf{Z}_{2j}-\bar{\mathbf{Z}}_{2j}\right)+\\&+ \sum_{j=1}^{n}\mathcal{P}\left(\mathbf{Z}_{j}+\bar{\mathbf{Z}}_{j}-\mathbf{Z}_{j-1}-\bar{\mathbf{Z}}_{j-1}\right)
\end{aligned}
\end{equation}
Similarly to the case of odd $n$, we gauge the isometries and set $\mathbf{Z}_{j}=\mathbf{S}_{j}=0$:
\begin{equation}
\begin{aligned}
&\mathcal{K} = {i\over 2}\sum_{j=1}^{\frac{n-2}{2}}\left(\mathbf{V}_{2j-1}-\mathbf{V}_{2j+1}\right)\left(\bar{\mathbf{X}}_{2j}-\mathbf{X}_{2j}\right) + \sum_{j=1}^{n} \mathcal{P}\left(\mathbf{V}_{j}-\mathbf{V}_{j-1}\right) \\&+\sum_{j=1}^{\frac{n-2}{2}}\left(\mathbf{Y}_{2j}\left(\mathbf{V}_{2j}+\mathbf{X}_{2j}\right)+\bar{\mathbf{Y}}_{2j}\left(\mathbf{V}_{2j}+\bar{\mathbf{X}}_{2j}\right)\right) + \sum_{j=1}^{\frac{n}{2}}\mathbf{V}_{2j-1}\left(\mathbf{\Sigma}_{2j-1}+\bar{\mathbf{\Sigma}}_{2j-1}\right)
\end{aligned}
\end{equation}

One can notice that in this case there are not enough semi-chiral fields $\mathbf{X}$ to eliminate all of the twisted chiral fields $\mathbf{\Sigma}$. However, shifting $\mathbf{X}_{2m}\mapsto 2i\left(\mathbf{X}_{2m}-\sum_{i=1}^{m}\mathbf{\Sigma}_{2i-1}\right)$ one can collect all of the $\mathbf{\Sigma}$'s as a factor in front of  $\mathbf{V}_{n-1}$ (here $\mathbf{X}_{n}\equiv \mathbf{X}_{0} \equiv 0$ and  $\mathbf{Y}_{2j-1}=\mathbf{X}_{2j}-\mathbf{X}_{2j-2}$):
\begin{equation}
\begin{aligned}
    &\mathcal{K} = \sum_{j=1}^{n-2}\mathbf{V}_{j}\mathbf{w}_{j} + \mathbf{V}_{n-1}\left(-\mathbf{X}_{n-2}-\bar{\mathbf{X}}_{n-2}+\sum_{i=1}^{n/2}\left(\mathbf{\Sigma}_{2i-1}+\bar{\mathbf{\Sigma}}_{2i-1}\right)\right) \\&+\sum_{j=1}^{n} \mathcal{P}\left(\mathbf{V}_{j}-\mathbf{V}_{j-1}\right) +2i\sum_{j=1}^{\frac{n-2}{2}}\left(\mathbf{Y}_{2j}\mathbf{X}_{2j}-\bar{\mathbf{Y}}_{2j}\bar{\mathbf{X}}_{2j}\right)
\end{aligned}
\end{equation}

Let us label the remaining twisted chiral field as $\mathbf{\Sigma}:=\sum_{i=1}^{n/2}\mathbf{\Sigma}_{2i-1}$. As a result, the system  contains one twisted chiral field $\mathbf{\Sigma}$  and $n-2$ semi-chiral  multiplets ($\mathbf{X}$ and $\mathbf{Y}$). Denoting  $\mathbf{w}_{n-1}=\mathbf{Y}_{n-1}+\bar{\mathbf{Y}}_{n-1}+\mathbf{\Sigma}+\bar{\mathbf{\Sigma}}$, we obtain the final expression for the generalized K\"ahler potential:
\begin{equation}
    \mathcal{K} = \sum_{j=1}^{n-1}\mathbf{V}_{j}\mathbf{w}_{j}+\sum_{j=1}^{n} \mathcal{P}\left(\mathbf{V}_{j}-\mathbf{V}_{j-1}\right) +2i\sum_{j=1}^{\frac{n-2}{2}}\left(\mathbf{Y}_{2j}\mathbf{X}_{2j}-\bar{\mathbf{Y}}_{2j}\bar{\mathbf{X}}_{2j}\right)\,.
\end{equation}

\vspace{0.3cm}    
    \setstretch{0.8}
    \setlength\bibitemsep{5pt}
    \printbibliography

\end{document}